\documentclass[10pt]{article}

\usepackage{pifont}
\usepackage{amsmath}
\usepackage{cases}
\usepackage{amsfonts}
\usepackage{latexsym}
\usepackage{amssymb}
\usepackage{stmaryrd}
\usepackage{overpic}
\usepackage{color}
\usepackage{abstract}
\usepackage{indentfirst}
\usepackage{booktabs}
\usepackage{url}
\usepackage{verbatim}
\usepackage{diagbox}
\usepackage{multirow} 
\usepackage{graphicx}  
\usepackage{float}  
\usepackage{subfigure} 
\usepackage{subfig}
\usepackage{bm}
\usepackage[numbers,sort&compress]{natbib}
\usepackage{graphicx}
\usepackage{epstopdf}

\makeatletter

\newcommand{\Rmnum}[1]{\expandafter\@slowromancap\romannumeral #1@}
\makeatother

\DeclareFontEncoding{FMS}{}{}
\DeclareFontSubstitution{FMS}{futm}{m}{n}
\DeclareFontEncoding{FMX}{}{}
\DeclareFontSubstitution{FMX}{futm}{m}{n}
\DeclareSymbolFont{fouriersymbols}{FMS}{futm}{m}{n}
\DeclareSymbolFont{fourierlargesymbols}{FMX}{futm}{m}{n}
\DeclareMathDelimiter{\VERT}{\mathord}{fouriersymbols}{152}{fourierlargesymbols}{147}

\newtheorem{theorem}{Theorem}[section]

\newtheorem{example}[theorem]{Example}
\newtheorem{lemma}[theorem]{Lemma}

\newtheorem{remark}{Remark}

\begin{document}

\begin{center}
\Large\bf Structure-preserving Randomized Neural Networks for\\ Incompressible Magnetohydrodynamics Equations
\end{center}

\vspace{2mm}

\begin{center}
 {\large\sc Yunlong Li}\footnote{School of Mathematics and Statistics, Xi'an Jiaotong University, Xi'an, Shaanxi 710049, P.R. China. E-mail: {\tt 4122107033@stu.xjtu.edu.cn}},\quad
{\large\sc Fei Wang}\footnote{School of Mathematics and Statistics \& State Key Laboratory of Multiphase Flow in Power Engineering, Xi'an Jiaotong University, Xi'an, Shaanxi 710049, China. The work of this author was partially supported by the Major Research Plan of
			the National Natural Science Foundation of China (Grant No.\ 92470115) and Tianyuan Fund for Mathematics of the National Natural Science Foundation of China (Grant No. 12426105). Email: {\tt feiwang.xjtu@xjtu.edu.cn}}, \quad
{\large\sc Lingxiao Li}\footnote{School of Mathematics and Statistics, Henan University, Kaifeng, Henan 475004, China. Center for Applied Mathematics of Henan Province, Henan University, Zhengzhou, Henan 450046, China. The work of this author was partially supported by National Natural Science Foundation of China (Grant No.\ 11901042).  E-mail: {\tt lilingxiao@lsec.cc.ac.cn}}

\end{center}

\vspace{2mm}

\begin{quote} 
\noindent{}{\bf Abstract}: 
The incompressible magnetohydrodynamic (MHD) equations are fundamental in many scientific and engineering applications. However, their strong nonlinearity and dual divergence-free constraints make them highly challenging for conventional numerical solvers. To overcome these difficulties, we propose a Structure-Preserving Randomized Neural Network (SP-RaNN) that automatically and exactly satisfies the divergence-free conditions. Unlike deep neural network (DNN) approaches that rely on expensive nonlinear and nonconvex optimization, SP-RaNN reformulates the training process into a linear least-squares system, thereby eliminating nonconvex optimization. The method linearizes the governing equations through Picard or Newton iterations, discretizes them at collocation points within the domain and on the boundaries using finite-difference schemes, and solves the resulting linear system via a linear least‑squares procedure. By design, SP-RaNN preserves the intrinsic mathematical structure of the equations within a unified space–time framework, ensuring both stability and accuracy. Numerical experiments on the Navier–Stokes, Maxwell, and MHD equations demonstrate that SP-RaNN achieves higher accuracy, faster convergence, and exact enforcement of divergence-free constraints compared with both traditional numerical methods and DNN-based approaches. This structure-preserving framework provides an efficient and reliable tool for solving complex PDE systems while rigorously maintaining their underlying physical laws.


\end{quote}

\vspace{2mm}

\section{Introduction}

The magnetohydrodynamics (MHD) equations govern the dynamic interplay between electrically conducting fluids and electromagnetic fields, with wide-ranging applications across scientific and engineering domains, such as metallurgical processes, magnetic confinement fusion, and liquid metal pumps (\cite{Davidson2001,Gerbeau2006,Muller2001,Jardin2010}). MHD represents a complex multiphysics system in which the movement of the conducting fluid generates electric currents that, in turn, modify the surrounding electromagnetic field. The interaction between the current and magnetic field produces the Lorentz force, driving the fluid in a direction perpendicular to both the current and the magnetic field.

This paper focuses on the incompressible MHD problem, which is characterized by a coupled system of Navier–Stokes and Maxwell equations. Extensive research has been conducted on finite element methods (FEM) for solving incompressible MHD equations. Studies such as \cite{Gerbeau2000,Greif2010,Zhang2014,Xie2024,Schotzau2004} focus on steady-state models, while others (\cite{Gao2019,Dong2018,Zhang2015}) address unsteady models. The incompressible MHD equations impose two essential divergence-free constraints on the velocity and magnetic fields, corresponding to mass and magnetic flux conservation. Ensuring that these constraints are exactly satisfied is crucial to avoid instabilities in numerical methods (\cite{Toth2000,Linke2009,John2017,Jiang1996,Linke2014}). Various divergence-free FEM techniques have been proposed, including for incompressible flow (\cite{Cockburn2007,Xu2010}), Maxwell equations (\cite{Cockburn2004,Huang2012}), and incompressible MHD equations (\cite{Zheng2018, Zheng2021, Xie2024, Hu2015, Hu2017}).

In recent years, neural networks (NNs) have emerged as powerful tools for solving partial differential equations (PDEs), due to their universal approximation capability (\cite{Barron1993,Chen1995}). Several neural network-based methods have been developed, including the Deep Ritz method (\cite{E2017}), the Deep Galerkin method (\cite{Sirignano2019}), and the Physics-Informed Neural Networks (\cite{George2019}). The total error in NN-based methods typically consists of approximation, statistical, and optimization errors, with optimization error being the dominant factor. Since training NNs involves solving non-convex, nonlinear optimization problems, the process can be computationally intensive and prone to local minima, limiting the accuracy and efficiency of current NN-based approaches compared to traditional numerical methods.

Randomized Neural Networks (RaNNs) represent a subset of NNs in which only the weights between the final hidden layer and the output layer are adjusted during training, transforming the problem into a linear least-squares optimization (\cite{Pao1994,Huang2006,Huang2011}). This reduces the optimization error while maintaining competitive approximation accuracy. The approximation capabilities of RaNNs have been thoroughly investigated in \cite{Igelnik1995,Neufeld2023} and other references. RaNNs have shown promise for solving PDEs, as explored in recent studies. For instance, Dong et al. enhanced the approximation capabilities of extreme learning machines (a type of RaNN) through domain decomposition, enforcing continuity across shared boundaries (\cite{Dong2021,Dong2023}), while Chen et al. extended this approach to overlapping domain decomposition (\cite{Chen2022}). Shang, Wang, and Sun integrated RaNNs with the Petrov-Galerkin method to solve both linear and nonlinear PDEs (\cite{Shang2023,Shang2024}), and Sun, Dong, and Wang coupled local RaNNs using discontinuous Galerkin (DG) formulations to improve approximation quality (\cite{Sun2024,Sun2024-2}). Dang and Wang combined RaNNs with the hybridized discontinuous Petrov-Galerkin method for Stokes-Darcy problems (\cite{Dang2023}), while Li and Wang incorporated RaNNs with the finite difference method for interface problems, coupling local RaNNs via interface conditions (\cite{Li2023}). These studies highlight the efficiency of RaNNs in achieving high accuracy with lower computational costs, making them particularly effective for time-dependent and high-dimensional problems.

In this paper, we propose a novel RaNN-based method for solving MHD equations, offering a strategy that automatically satisfies divergence-free conditions, which reflects the fundamental physical law of conservation of magnetic flux

Existing NN-based approaches (\cite{George2021,Shang2024}) typically enforce these conditions at discrete points, which results in insufficient precision. Our proposed method ensures exact, automatic satisfaction of the divergence-free condition by leveraging the properties of randomized neural networks. We term this approach the Structure-Preserving Randomized Neural Network (SP-RaNN). The key concept is constructing divergence-free basis functions using equivalence conditions, allowing the RaNN to inherently satisfy the divergence-free constraint due to the properties of its function space.

Given the nonlinear nature of MHD equations, two common approaches exist to address the complexity: nonlinear least-squares computations during training (\cite{Dong2021}) or nonlinear iterations to linearize the equations (\cite{Shang2023}). This paper adopts the latter approach. SP-RaNN improves accuracy, reduces computational costs, and performs effectively in high Reynolds number scenarios. The method is applied to solve Navier-Stokes, Maxwell, and MHD equations, demonstrating its robustness and efficiency through comparisons with traditional NN methods and FEM approaches.

The structure of the paper is as follows: Section \ref{sec2} introduces the Navier-Stokes, Maxwell, and incompressible MHD equations. In Section \ref{sec3}, we describe the randomized neural network (RaNN) and demonstrate how to construct divergence-free RaNNs. Section \ref{sec:DFRaNNFDM} presents the SP-RaNN for solving MHD equations. Numerical results for the Stokes, Navier-Stokes, Maxwell, and MHD equations are presented in Section \ref{sec:NE}, showcasing the method’s accuracy and robustness. Finally, we conclude the paper with a summary and future research directions.

\section{Incompressible Magnetohydrodynamics Equations}
\label{sec2}

In this section, we present the equations considered in this study.

The incompressible magnetohydrodynamics (MHD) equations describe the coupling between the Navier-Stokes (NS) equations and Maxwell’s equations. First, we introduce the unsteady Navier-Stokes equations as follows:
\begin{equation} \label{N-S-model}
\left\{
\begin{array}{rrll}
    \frac{\partial \mathbf{u}}{\partial t} + (\mathbf{u}\cdot \nabla)\mathbf{u} + \nabla p - \frac{1}{{\rm Re}}\Delta \mathbf{u} &=& \mathbf{f},\quad& {\rm in}\ \Omega \times I, \\
    \nabla \cdot \mathbf{u} &=& 0,\quad& {\rm in}\ \Omega \times I, \\
    \mathbf{u} &=& \mathbf{g}_{D},\quad& {\rm on}\ \Gamma_D \times I, \\
    \frac{\partial \mathbf{u}}{\partial \mathbf{n}} &=& \mathbf{g}_{N},\quad& {\rm on}\ \Gamma_N \times I, \\
    \mathbf{u} &=& \mathbf{u}_0, \quad& {\rm on}\ \Omega \times \{0\},
\end{array}
\right.
\end{equation}
where $\Omega$ is the spatial domain, $I$ is the time interval, $\mathbf{f}$ is an external force, and $t$, $\mathbf{u}$, and $p$ represent the non-dimensional time, velocity, and pressure, respectively. Here, Re denotes the Reynolds number. The first two equations in \eqref{N-S-model} correspond to the conservation of momentum and mass. The divergence-free condition \( \nabla \cdot \mathbf{u} = 0 \) is essential for ensuring mass conservation, physical consistency, and numerical stability in incompressible Navier-Stokes simulations. Violating this condition can result in unphysical flow patterns, pressure artifacts, and instability. Effective enforcement of \( \nabla \cdot \mathbf{u} = 0 \) is a cornerstone of robust and accurate numerical methods for solving incompressible flow problems.

Next, we examine Maxwell’s equations in a simple homogeneous medium under perfect conductivity conditions:
\begin{equation} \label{maxwell-model}
\left\{
\begin{array}{rrll}
    \mu \frac{\partial \mathbf{H}}{\partial t} &=& -\nabla \times \mathbf{E},\quad& {\rm in}\ \Omega \times I, \\
    \epsilon \frac{\partial \mathbf{E}}{\partial t} &=& \nabla \times \mathbf{H},\quad& {\rm in}\ \Omega \times I, \\
    \nabla \cdot \mathbf{H} &=& 0,\quad& {\rm in}\ \Omega \times I, \\
    \mathbf{n} \times \mathbf{E} &=& 0,\quad  \mathbf{H} \cdot \mathbf{n}\;\;\; =\;\;\; 0,& {\rm on}\ \partial \Omega \times I, \\
    \mathbf{E} &=& \mathbf{E}_0, \quad  \mathbf{H}\;\;\; =\;\;\; \mathbf{H}_0, & {\rm on}\ \Omega \times \{0\}, \\
\end{array}
\right.
\end{equation}
where $\mathbf{H}$ denotes the magnetic field, $\mathbf{E}$ represents the electric field, $\mu$ is the magnetic permeability, and $\epsilon$ stands for electric permittivity. The first three equations in \eqref{maxwell-model} correspond to Faraday’s law of electromagnetic induction, the Amp\'ere–Maxwell law, and Gauss’s law for magnetism, respectively. The divergence-free condition \( \nabla \cdot \mathbf{H} = 0 \) encapsulates the fundamental physical principle of magnetic flux conservation, signifying that magnetic field lines neither originate nor terminate within a medium—they either form closed loops or extend to infinity. This condition is crucial for accurately capturing the physical and mathematical characteristics of Maxwell’s equations. Ensuring \( \nabla \cdot \mathbf{H} = 0 \) is vital for maintaining stability, physical realism, and reliability in numerical simulations, making it a foundational element of any robust computational framework for solving Maxwell’s equations.

Finally, we present the unsteady MHD equations:
\begin{equation} \label{MHD-model2}
\left\{
\begin{array}{rrll}
    \frac{\partial \mathbf{u}}{\partial t} + (\mathbf{u}\cdot \nabla)\mathbf{u} + \nabla p - \frac{1}{{\rm Re}}\Delta \mathbf{u} - \frac{1}{\mu}(\nabla \times \mathbf{B})\times\mathbf{B} &=& \mathbf{f},\quad& {\rm in}\ \Omega \times I, \\
    \frac{\partial \mathbf{B}}{\partial t} + \frac{1}{\sigma \mu}\nabla \times(\nabla \times \mathbf{B}) - \nabla \times(\mathbf{u} \times \mathbf{B}) &=& \mathbf{g},\quad& {\rm in}\ \Omega \times I, \\
    \nabla \cdot \mathbf{u}\;\;\; =\;\;\; 0,\quad  \nabla \cdot \mathbf{B} &=& 0, & {\rm in}\ \Omega \times I, \\
    \mathbf{u} \;\;\;=\;\;\; \mathbf{u}_{D},\quad  \mathbf{B} \times \mathbf{n} &=& \mathbf{B}_{D}, & {\rm on}\ \partial \Omega \times I, \\
    \mathbf{u}\;\;\; = \;\;\; \mathbf{u}_0,\quad  \mathbf{B} &=& \mathbf{B}_0, & {\rm on}\ \Omega \times \{0\}, \\
\end{array}
\right.
\end{equation}
In this system, $ \mathbf{u} $ is the velocity vector, $ p $ is the pressure, $ \mathbf{B} $ denotes the magnetic field, $ \mu $ is the magnetic permeability, and $ \sigma $ is the electrical conductivity. The first two equations represent the interaction between the fluid flow and the magnetic field, while the third and fourth equations describe the conservation of mass and magnetic flux, respectively.

\section{Structure-Preserving Randomized Neural Networks Method}\label{sec3}
\subsection{Randomized Neural Networks}
\label{sec:RaNN}

\begin{figure*}[!htb]
\centering
    \subfigure[Fully connected neural network]{\includegraphics[width=0.45\textwidth]{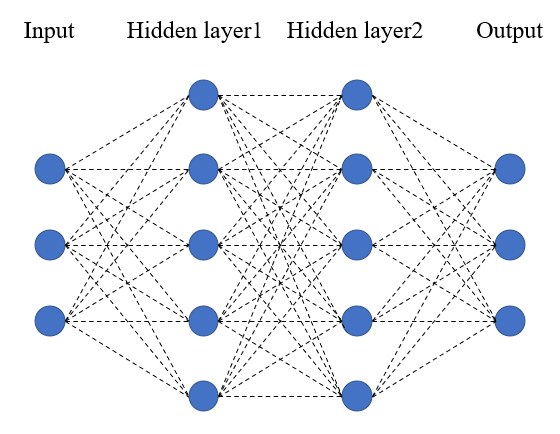}}\quad
    \subfigure[Randomized neural network]{\includegraphics[width=0.45\textwidth]{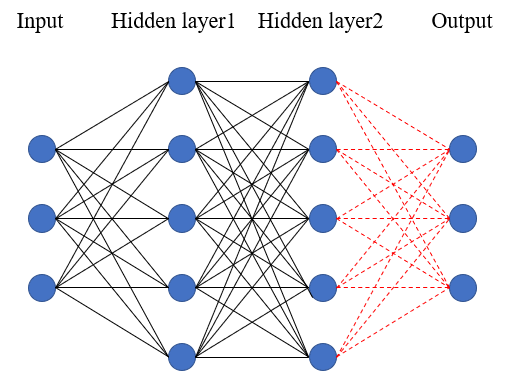}}
    \caption{The structure of two-hidden-layers neural networks: the dotted line representing weights and bias which are tunable, and the solid line representing weights and bias which are randomly given and fixed.}
    \label{NN}
    \vspace{0.1in}
\end{figure*}

First, we introduce the fully connected neural network $\Psi: \mathbb{R}^{n_0} \rightarrow \mathbb{R}^{n_D}$ (see Figure \ref{NN}(a)) with depth $\rm D$. The width of each layer is denoted by $n_l$, for $l = 0, \dots, D$. The network is defined as follows:
\begin{align*}
    &\Psi_{0}(\mathbf{x}) = \mathbf{x},\\
    &\Psi_{k}(\mathbf{x}) = \rho(W_{k}\Psi_{k-1}+b_{k}), \quad k=1,...,D-1,\\
    &\Psi = \Psi_{D} = W_{D}\Psi_{D-1},
\end{align*}
where $\rho$ is the activation function, and $W_{k} \in \mathbb{R}^{n_{k} \times n_{k-1}}$ and $b_{k} \in \mathbb{R}^{n_k}$ are the weights and biases for the $k$-th layer, determined through training.

In a randomized neural network (RaNN), the weights $W_1, \dots, W_{D-1}$ and biases $b_1, \dots, b_{D-1}$ are generated randomly and fixed. Only the weights $W_D$ between the last hidden layer and the output layer are adjusted, resulting in the RaNN (Figure \ref{NN}(b)), denoted as $\mathbf{u}_{\rho}$. The neurons $\{\psi_1, \dots, \psi_{n_{D-1}}\}$ in the $(D-1)$-th layer form a set of basis functions, i.e., $u_{\rho} \in \mathcal{N}_{\rho}^D := {\rm span}\{\psi_1, \dots, \psi_{n_{D-1}} \}$, where $u_{\rho}$ is a component of $\mathbf{u}_{\rho}$. Assuming $\mathbf{u}_{\rho} = (u^1_{\rho}, u^2_{\rho}, u^3_{\rho})^{\rm T}$, i.e., $n_{D} = 3$, the components can be expressed as a linear combination of the basis functions:
\begin{equation}\label{linear_combination1}
\begin{aligned}
    u^1_{\rho} = \sum\limits_{i=1}^{n_{D-1}} \alpha_{i} \psi_{i},\quad u^2_{\rho} = \sum\limits_{i=1}^{n_{D-1}} \beta_{i} \psi_{i},\quad 
    u^3_{\rho} = \sum\limits_{i=1}^{n_{D-1}} \gamma_{i} \psi_{i}.
\end{aligned}
\end{equation}
Thus, training the RaNN involves determining the values of $\{\alpha_i\}_{i=1}^{n_{D-1}}, \{\beta_i\}_{i=1}^{n_{D-1}}$, and $\{\gamma_i\}_{i=1}^{n_{D-1}}$, which can be solved using a linear optimization problem rather than a non-convex and nonlinear optimization. The least-squares method is an efficient approach for this.

\subsection{Structure-Preserving RaNNs}
\label{sec:SP-RaNN}

In \cite{John2017}, various interpretations of a function being ``divergence-free" were summarized, ranging from the strongest to the weakest: pointwise divergence-free, weakly divergence-free, and discretely divergence-free. In this work,  we develop a Structure-Preserving Randomized Neural Network (SP-RaNN) to achieve pointwise divergence-free solutions. The key innovation lies in the construction of divergence-free neural network basis functions, ensuring that the network inherently satisfies the pointwise divergence-free condition without requiring additional constraints or corrections. By preserving this essential structure, the SP-RaNN provides a mathematically consistent and physically accurate framework for solving problems where divergence-free conditions are critical.

To construct a divergence-free neural network, we recall the following result.
\begin{lemma}[\cite{Pozrikidis1997}]
 In a simply connected region, for a vector-valued function $\mathbf{u}:\mathbb{R}^3 \rightarrow\mathbb{R}^3$, the following holds:
\begin{align} \label{df-3d}
    \nabla \cdot \mathbf{u}=0 &\iff \exists\ \bm{\Psi}:\mathbb{R}^3 \rightarrow\mathbb{R}^3 \ s.t.\ \mathbf{u}=\nabla \times \bm{\Psi}.
\end{align}
\end{lemma}

For the two-dimensional case, $\mathbf{u}:\mathbb{R}^2 \rightarrow\mathbb{R}^2$, we have a similar result:
\begin{align}\label{df-2d}
    \nabla \cdot \mathbf{u}=0 \iff \exists\ \psi:\mathbb{R}^2 \rightarrow\mathbb{R}\ s.t.\ \mathbf{u}={\rm curl} \psi. 
\end{align}
where
\begin{align*}
    {\rm curl} \psi= \left(\frac{\partial \psi}{\partial y}, -\frac{\partial \psi}{\partial x}\right)^{\rm T}.
\end{align*}

Using these results, we construct SP-RaNNs. Given any vector-valued function $\bm{\Psi}(x,y,z)=(\psi_1, \psi_2, \psi_3)$, from \eqref{df-3d}, we know that $\bm{\phi} = (\phi_1, \phi_2, \phi_3)$ is divergence-free, where
\begin{align*}
    \phi_{1}(\mathbf{x})=\frac{\partial \psi_{3}}{\partial y} - \frac{\partial \psi_{2}}{\partial z},\quad 
    \phi_{2}(\mathbf{x})=\frac{\partial \psi_{1}}{\partial z} - \frac{\partial \psi_{3}}{\partial x} ,\quad 
    \phi_{3}(\mathbf{x})=\frac{\partial \psi_{2}}{\partial x} - \frac{\partial \psi_{1}}{\partial y}.
\end{align*}
Consider a neural network where $\psi_i = \psi_i(w_{i1}x+w_{i2}y+w_{i3}z+b_i)=\psi_i(h_i(\mathbf{x}))$ for $i=1,2,3$. Then we have
\begin{equation}\label{basis-function-3dr}
\begin{aligned}
    \phi_{1}(\mathbf{x})&=w_{32}\psi_{3}'(h_{3}(\mathbf{x}))-w_{23}\psi_{2}'(h_{2}(\mathbf{x})),\\
    \phi_{2}(\mathbf{x})&=w_{13}\psi_{1}'(h_{1}(\mathbf{x}))-w_{31}\psi_{3}'(h_{3}(\mathbf{x})),\\
    \phi_{3}(\mathbf{x})&=w_{21}\psi_{2}'(h_{2}(\mathbf{x}))-w_{12}\psi_{1}'(h_{1}(\mathbf{x})).
\end{aligned}
\end{equation}
By choosing $\psi_{i}' = \rho_i$, the RaNN function
\begin{align*}
    &\mathbf{u}_{\rho}=\sum\limits_{j=1}^{m} \alpha_{j} \bm{\phi}^{j}
\end{align*}
is divergence-free, where
\begin{align*}
    &\bm{\phi}^{j}=(\phi_{1}^{j},\phi_{2}^{j},\phi_{3}^{j})^T,\\
    &\phi_{1}^j(\mathbf{x})=w^{j}_{32}\rho_3^j(h^{j}_{3}(\mathbf{x}))-w^{j}_{23}\rho_2^j(h^{j}_{2}(\mathbf{x})),\\
    &\phi_{2}^j(\mathbf{x})=w^{j}_{13}\rho_1^j(h^{j}_{1}(\mathbf{x}))-w^{j}_{31}\rho_3^j(h^{j}_{3}(\mathbf{x})),\\
    &\phi_{3}^j(\mathbf{x})=w^{j}_{21}\rho_2^j(h^{j}_{2}(\mathbf{x}))-w^{j}_{12}\rho_1^j(h^{j}_{1}(\mathbf{x})),\\
    &h^{j}_i(\mathbf{x})=w^{j}_{i1}x+w^{j}_{i2}y+w^{j}_{i3}z+b_i,\quad {\rm for}\ j=1,\cdots,m.
\end{align*}

\begin{figure*}[!htb]
\centering
    \subfigure[2D divergence-free RaNN]{\includegraphics[width=0.45\textwidth]{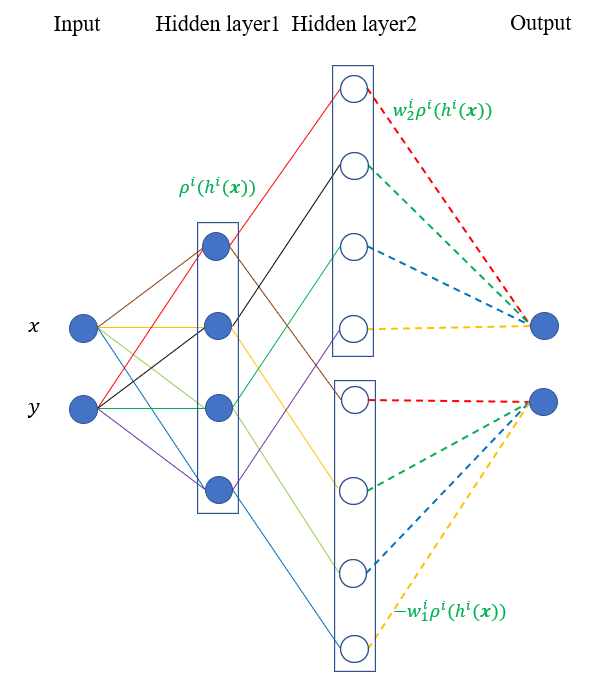}}\quad
    \subfigure[3D divergence-free RaNN]{\includegraphics[width=0.45\textwidth]{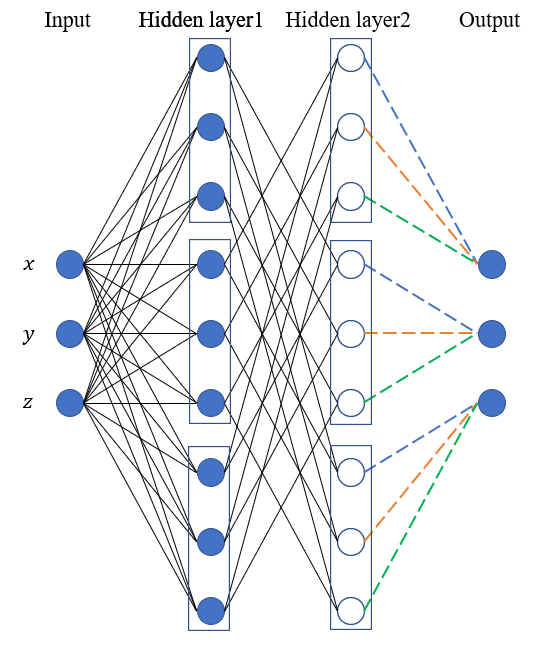}}
    \caption{The structure of a two-hidden-layer neural network: the dotted lines represent tunable weights and biases, while the solid lines represent weights and biases that are randomly initialized and fixed.
    }
    \label{RaNN}
    \vspace{0.1in}
\end{figure*}

For the two-dimensional case, given any scalar-valued function $\psi(x,y)$, the vector $\bm{\Phi}=(\phi_1,\phi_2)$ is divergence-free, where 
$$
\phi_1(\mathbf{x})=\frac{\partial \psi}{\partial y},\ \phi_2(\mathbf{x})=-\frac{\partial \psi}{\partial x}.
$$
Considering a neural network with $\psi=\psi(w_1x+w_2y+b)=\psi(h(\mathbf{x}))$, we obtain
\begin{equation}\label{basis-function-2dr}
    \begin{aligned}
        \phi_{1}(\mathbf{x})&=w_2\psi'(h(\mathbf{x})),\\
        \phi_{2}(\mathbf{x})&=-w_1\psi'(h(\mathbf{x})).
    \end{aligned}
\end{equation}
Similarly, by choosing $\psi'=\rho$, the RaNN function
\begin{align*}
    &\mathbf{u}_{\rho}=\sum\limits_{j=1}^{m} \gamma_{j} \bm{\phi}^{j}
\end{align*}
is divergence-free, where
\begin{align*}
    &\bm{\phi}^{j}=(\phi_{1}^{j},\phi_{2}^{j})^T,\\
    &\phi_{1}^j(\mathbf{x})=w^{j}_{2}\rho^j(h^{j}(\mathbf{x})),\\
    &\phi_{2}^j(\mathbf{x})=-w^{j}_{1}\rho^j(h^{j}(\mathbf{x})),\\
    &h^{j}(\mathbf{x})=w^{j}_{1}x+w^{j}_{2}y+b_i,\quad {\rm for}\ j=1,\cdots,m.
\end{align*}

The structures of the 2D and 3D divergence-free RaNNs are shown in Figure \ref{RaNN}. The weights between the input and hidden layer1 are randomly chosen and fixed, while those between the hidden layer1 and hidden layer2 are determined accordingly. In Figure \ref{RaNN} (a), corresponding weights (solid lines of the same color) are equal. The same holds in Figure \ref{RaNN} (b), though some details are omitted for clarity. The bias in the second hidden layer is zero, and the activation functions in this layer are identity maps. The weights connecting the second hidden layer to the output layer can be determined through a least-squares method, with dotted lines of the same color representing equal weights.

\begin{remark}
    Previously, we assume that the number of neurons in the input layer is 2 and 3 for the 2D and 3D cases, respectively. However, for time-dependent problems, this number may increase to 3 and 4, respectively. Since the divergence-free property is independent of time, the strategies for constructing the structure of SP-RaNNs remain applicable.
\end{remark}

\begin{remark}
    In numerical experiments, we choose all $\rho_i^j$ to be the same function for the 3D case and all $\rho^j$ to be the same for the 2D case. If certain features of the solution are known, different $\rho_i^j$ or $\rho^j$ could be selected to better approximate the solution.
\end{remark}

\section{SP-RaNN for MHD Equations}
\label{sec:DFRaNNFDM}

In this section, we demonstrate how the SP-RaNN method can be applied to solve the 3D MHD equations. First, we linearize the equations \eqref{MHD-model2} using nonlinear iteration methods. This paper focuses on the Picard and Newton iteration methods, with the formulations as follows:
\begin{equation} \label{Picard2}
    \left\{
    \begin{array}{l}
        \frac{\partial \mathbf{u}^{n+1}}{\partial t} + (\mathbf{u}^{n}\cdot \nabla)\mathbf{u}^{n+1} + \nabla p^{n+1} - \frac{1}{{\rm Re}}\Delta \mathbf{u}^{n+1} - \frac{1}{\mu}(\nabla \times \mathbf{B}^{n+1})\times\mathbf{B}^{n} = \mathbf{f},\\
        \frac{\partial \mathbf{B}^{n+1}}{\partial t} + \frac{1}{\sigma \mu}\nabla \times(\nabla \times \mathbf{B}^{n+1}) - \big(\mathbf{u}^{n}(\nabla \cdot \mathbf{B}^{n+1})-(\mathbf{u}^{n}\cdot \nabla)\mathbf{B}^{n+1}-\mathbf{B}^{n}(\nabla \cdot \mathbf{u}^{n+1})+(\mathbf{B}^{n}\cdot \nabla)\mathbf{u}^{n+1}\big) = \mathbf{g}, \\
        \nabla \cdot \mathbf{u}^{n+1} = 0,\ \nabla \cdot \mathbf{B}^{n+1} = 0, \\
        \mathbf{u}^{n+1} = \mathbf{u}_{D},\ \mathbf{B}^{n+1} \times \mathbf{n} = \mathbf{B}_{D},\qquad {\rm on}\ \partial \Omega \times I, \\
        \mathbf{u}^{n+1} = \mathbf{u}_0,\ \mathbf{B}^{n+1} = \mathbf{B}_0,\qquad {\rm on}\ \Omega \times \{0\}.
    \end{array}
    \right.
\end{equation}

\begin{equation} \label{Newton}
    \left\{
    \begin{array}{l}
        \frac{\partial \mathbf{u}^{n+1}}{\partial t} + ((\mathbf{u}^{n+1}\cdot \nabla)\mathbf{u}^{n} + (\mathbf{u}^{n}\cdot \nabla)\mathbf{u}^{n+1}) + \nabla p^{n+1} - \frac{1}{{\rm Re}}\Delta \mathbf{u}^{n+1} 
        - \frac{1}{\mu}((\nabla \times \mathbf{B}^{n+1})\times\mathbf{B}^{n} \\ 
   \qquad\qquad     +(\nabla \times \mathbf{B}^{n})\times\mathbf{B}^{n+1}) = \mathbf{f} + (\mathbf{u}^{n}\cdot \nabla)\mathbf{u}^{n}-\frac{1}{\mu}(\nabla \times \mathbf{B}^{n})\times\mathbf{B}^{n},\\
        \frac{\partial \mathbf{B}^{n+1}}{\partial t} + \frac{1}{\sigma \mu}\nabla \times(\nabla \times \mathbf{B}^{n+1}) - 
        \big((\mathbf{u}^{n}(\nabla \cdot \mathbf{B}^{n+1})+\mathbf{u}^{n+1}(\nabla \cdot \mathbf{B}^{n})\big)
        -\big((\mathbf{u}^{n+1}\cdot \nabla)\mathbf{B}^{n}+(\mathbf{u}^{n}\cdot \nabla)\mathbf{B}^{n+1}\big)\\
     \qquad\qquad    - (\mathbf{B}^{n}(\nabla \cdot \mathbf{u}^{n+1})+\mathbf{B}^{n+1}(\nabla \cdot \mathbf{u}^{n}))+
        \big((\mathbf{B}^{n+1}\cdot \nabla)\mathbf{u}^{n}+(\mathbf{B}^{n}\cdot \nabla)\mathbf{u}^{n+1})\big) \\
    \qquad\qquad      = \mathbf{g}-(\mathbf{u}^{n}(\nabla \cdot \mathbf{B}^{n})-(\mathbf{u}^{n}\cdot \nabla)\mathbf{B}^{n}-\mathbf{B}^{n}(\nabla \cdot \mathbf{u}^{n})+(\mathbf{B}^{n}\cdot \nabla)\mathbf{u}^{n}), \\
        \nabla \cdot \mathbf{u}^{n+1} = 0,\ \nabla \cdot \mathbf{B}^{n+1} = 0, \\
        \mathbf{u}^{n+1} = \mathbf{u}_{D},\ \mathbf{B}^{n+1} \times \mathbf{n} = \mathbf{B}_{D}, \qquad  {\rm on}\ \partial \Omega \times I, \\
        \mathbf{u}^{n+1} = \mathbf{u}_0,\ \mathbf{B}^{n+1} = \mathbf{B}_0, \qquad  {\rm on}\ \Omega \times \{0\}.
    \end{array}
    \right.
\end{equation}

In implementing the RaNN method, we adopt a space-time approach, treating time and spatial variables equivalently. This eliminates the need for time-stepping iterations and reduces error accumulation. In the space-time domain, the initial condition is treated as a boundary condition. We use two SP-RaNNs to approximate $\mathbf{u}$ and $\mathbf{B}$ separately, and an additional RaNN to approximate $p$. For simplicity, we assume all networks have the same number of basis functions. Based on the derivations in Section \ref{sec:SP-RaNN}, we define:
\begin{align}
    \mathbf{u}_{\rho} = \sum\limits_{i=1}^{m} \alpha^u_{i} \bm{\phi}^{u}_i(\mathbf{x}),\quad 
    \mathbf{B}_{\rho} = \sum\limits_{i=1}^{m} \alpha^B_{i} \bm{\phi}^{B}_i(\mathbf{x}),\quad 
    p_{\rho} = \sum\limits_{i=1}^{m} \alpha^p_{i} \phi^p_i(\mathbf{x}). \label{linear-express}
\end{align}

The total number of degrees of freedom is $3m$. By substituting $\mathbf{u}^{n+1}, \mathbf{B}^{n+1},$ and $p^{n+1}$ in \eqref{Picard2} (or \eqref{Newton}) with their corresponding expressions from \eqref{linear-express}, we discretize the equations at $N_c$ collocation points, which include $N_1$ interior points and $N_2$ boundary points. This yields a linear system for solving $X = (\alpha^u_1, \dots, \alpha^u_m, \alpha^B_1, \dots, \alpha^B_m, \alpha^p_1, \dots, \alpha^p_m)^T$:
\begin{equation}
    \begin{aligned}
        A^1X = F^1,\ A^2X = F^2,\ A^3X &= F^3,\\
        C^1X = G^1,\ C^2X = G^2,\ C^3X &= G^3,\\
        B^1X = b^1,\ B^2X = b^2,\ B^3X &= b^3,\\
        B^4X = b^4,\ B^5X = b^5,\ B^6X &= b^6,\\
        D^1X = d^1,\ D^2X = d^2,\ D^3X &= d^3,\\
        D^4X = d^4,\ D^5X = d^5,\ D^6X &= d^6,\\
    \end{aligned}
\end{equation}
where $A^i, C^i$ ($i = 1,2,3$) are $N_1 \times 3m$ matrices, $B^j$ and $D^j$ ($j = 1, \dots, 6$) are $\frac{N_2}{7} \times 3m$ matrices, corresponding to the first two equations in \eqref{Picard2}, boundary, and initial conditions. The matrix elements are derived from \eqref{Picard2} and \eqref{Newton}.

The sampling strategy is designed such that the ratio of sampling points aligns approximately with the measurement ratio. For instance, in the 3-dimensional case, we sample $N^2$ points on each boundary face and $N^3$ points within $\Omega$. Time-dependent problems are treated as $(d+1)$-dimensional problems, where initial conditions are regarded as boundary conditions. For further details, see \cite{Li2023}.

We use finite difference schemes to approximate derivatives in the linear system. Specifically, second-order central difference schemes are used:
\begin{align}
	\frac{\partial{u_\rho(x,y)}}{\partial x} &= \frac{u_\rho(x+h_1,y)-u_\rho(x-h_1,y)}{2h_1}, \label{diff_1}\\
	\frac{\partial^2{u_\rho(x,y)}}{\partial^2 x} &= \frac{u_\rho(x+h_2,y) -2u_\rho(x,y)  + u_\rho(x-h_2,y) }{h_2^2}. \label{diff_2}
\end{align}
The initial vector $X = (0, ..., 0)^T$ is used, and the least-squares method is employed to solve the linear system at each iteration step. The final $X$ obtained at the last iteration provides the weights for the SP-RaNNs and RaNN.

\begin{remark}\label{constrain_p}   
    To ensure the uniqueness of $p$, an additional condition $\int_{\Omega} p \, d\mathbf{x} = c$, where $c$ is a constant, must be imposed. There are two common approaches to enforce this constraint. One method is to shift both the approximate and exact pressures to align their means with a constant $c$ (\cite{George2021}). In this paper, we solve the integral equation directly using RaNNs. Specifically, Gauss-Legendre numerical integration is applied:
    \[
    \int_{\Omega} p \, d\mathbf{x} = \sum_{i=1}^{m} \alpha^p_i \int_{\Omega} \rho_i(\mathbf{x}) \, d\mathbf{x} = \sum_{i=1}^{m} \alpha^p_i \left( \sum_{j=1}^{n} \beta_j \rho_i(\mathbf{x}_j) \right),
    \]
    where $n$ is the number of Gauss-Legendre points, and $\beta_j$ are the integration weights. This method yields more accurate results. For steady-state problems, this adds an additional equation to the linear system. For unsteady problems, time slicing generates as many equations as there are time slices.
\end{remark}

\begin{remark}
The divergence-free conditions are crucial for accurately representing the physical and mathematical properties of MHD systems. Enforcing \( \nabla \cdot \mathbf{u} = 0 \) and \( \nabla \cdot \mathbf{B} = 0 \) ensures numerical stability, physical realism, and reliability, establishing these conditions as fundamental requirements for any robust computational framework for MHD equations. In the SP-RaNN method, these divergence-free conditions are inherently satisfied, as the basis functions constructed in Section \ref{sec:SP-RaNN} are designed to be divergence-free by construction. Consequently, explicit constraints such as \( \nabla \cdot \mathbf{u}^{n+1} = 0 \) and \( \nabla \cdot \mathbf{B}^{n+1} = 0 \) are unnecessary. Moreover, by using the same number of basis functions and collocation points, the SP-RaNN system is more computationally efficient than the standard RaNN, as it involves fewer rows and columns in the resulting linear system.
\end{remark}

\begin{remark}
To emphasize the boundary and initial conditions during training, we can multiply both sides of the corresponding equations ($B^i X = b^i$, $D^i X = d^i$, $i=1, \dots, 6$) by a positive constant $\gamma$, typically greater than 1. This ensures that the RaNN prioritizes boundary and initial conditions, which is crucial for solution accuracy, especially for Dirichlet boundary conditions. However, care must be taken to avoid overfitting, as excessively large $\gamma$ may reduce overall accuracy.
\end{remark}

\begin{remark}
   For long-time simulations, different temporal strategies can be adopted depending on the temporal regularity of the solution. 
When the time evolution of the solution is smooth, one may directly employ the space–time SP-RaNN formulation with a properly chosen network scale parameter $r$ (\cite{Dang2024}), which can effectively capture the coupled spatial–temporal behavior without introducing additional time-stepping errors. 

When the dynamics exhibit rapid or localized temporal variations, it is advisable to combine the SP-RaNN spatial approximation with a temporal discretization scheme. 
An adaptive time-stepping strategy may be adopted—using larger steps over smooth intervals and smaller steps near sharp transients—to achieve a balance between efficiency and accuracy. 
The parameter $r$ can also be adaptively updated at each time level according to the local temporal complexity or residual magnitude, thereby enhancing the network’s approximation capability in regions of fast temporal change. 
To further reduce computational cost, one can employ the linearly extrapolated schemes proposed in \cite{Zheng2018, Hou2015} to avoid nonlinear iterations within each time step.
    
\end{remark}

\section{Numerical Examples}
\label{sec:NE}

In this section, we apply the SP-RaNN to several problems, including the Stokes equations (\cite{Linke2014}), Navier-Stokes (N-S) equations (\cite{George2021}), Maxwell equations (\cite{Xie2013}), and MHD equations (\cite{Xie2024, Zheng2018}). The numerical results demonstrate the performance of the proposed approach. All algorithms are implemented in Python, and the least-squares problems are solved using the QR method provided by ``scipy.linalg.lstsq". Numerical experiments for SP-RaNN are conducted on a computer equipped with the 13th Gen Intel(R) Core(TM) i7-13700KF CPU @3.40GHZ processor and 32GB of RAM.

In the numerical experiments, unless explicitly specified, both RaNNs and SP-RaNNs use the activation function $\rho = \text{tanh}$. The finite difference step sizes are set to $h_1 = 10^{-6}$ and $h_2 = 5 \times 10^{-4}$ for the schemes in \eqref{diff_1} and \eqref{diff_2}, respectively. Initial weights and biases are sampled from a uniform distribution $U(-r, r)$. Throughout all examples, the number of basis functions in the RaNNs and SP-RaNNs is denoted by $m$, and the degrees of freedom (DoFs) refer to the total number of unknowns across both RaNN and SP-RaNN models.

To quantify errors over the domain $\Omega$, we use the $L^2$, semi-$H^1$, and $H^1$ norms, denoted as $\|\cdot\|_0$, $|\cdot|_1$, and $\|\cdot\|_1$, respectively. Additionally, for the space-time domain $\Omega \times I$, we define the corresponding space-time norms: $\|\cdot\|_0^{st}$ for the $L^2$ norm, $|\cdot|_1^{st}$ for the semi-$H^1$ norm, and $\|\cdot\|_1^{st}$ for the full $H^1$ norm. The relative $L^2$ error and space-time relative $L^2$ error for a given function $v$ are defined as follows:
\[
e_v = \frac{\Vert v - v_{\rho} \Vert_{0}}{\Vert v \Vert_{0}}, \qquad 
e^{st}_v = \frac{\Vert v - v_{\rho} \Vert_{0}^{st}}{\Vert v \Vert_{0}^{st}},
\]
where $v$ can be either a scalar-valued or vector-valued function. The errors are evaluated using Gauss-Legendre numerical integration. We use $20^d$ Gauss-Legendre quadrature points for Example \ref{ex_2d_ns} ($d=2$) and Example \ref{ex_2d_mhd} ($d=3$), and $10^d$ quadrature points for Example \ref{ex_3d_ns}, Example \ref{ex_maxwell}, and Example \ref{ex_3d_mhd} ($d=4$). All results are averaged over ten experiments.

\begin{example}[Steady Two-dimensional Stokes Equations]\label{ex_2d_ns}
In this example, we consider the steady two-dimensional Stokes equations:
\begin{equation} \label{ex-stokes-2d}
\left\{
\begin{array}{rrll}
    - \frac{1}{{\rm Re}}\Delta \mathbf{u} + \nabla p &=& \mathbf{f},\ {\rm in}\ \Omega, \\
    \nabla \cdot \mathbf{u} &=& 0,\ {\rm in}\ \Omega, \\
    \mathbf{u} &=& \mathbf{g}_{D},\ {\rm on}\ \partial \Omega,
\end{array}
\right.
\end{equation}
where Re is the Reynolds number. We test the same example from \cite{Linke2014} (Example 5.2), where $\Omega=(0,1)^2$, and the exact solution is given by:
\begin{equation}
    \left\{
    \begin{array}{rrll}
        u &=& x^2y^2(1 - x)^2(2y - 2) + 2x^2y(1 - x)^2(1 - y)^2 ,  \\
        v &=& -x^2y^2(1 - y)^2(2x - 2) - 2xy^2(1 - x)^2(1 - y)^2, \\
        p &=& x^3 + y^3 - \frac{1}{2}. \nonumber
    \end{array}
    \right.
    \end{equation}
\end{example}

We use both RaNN and SP-RaNN (using the same parameters) to approximate $\mathbf{u}$ for this problem. In RaNN, two RaNNs are used to approximate $\mathbf{u}$ and $p$ separately, while in SP-RaNN, a SP-RaNN approximates $\mathbf{u}$, and an RaNN approximates $p$. We set $r_1 = 1$ for the SP-RaNN (and RaNN) used to approximate $\mathbf{u}$, $r_2 = 0.5$ for the RaNN approximating $p$, and $N_c = 5180$ ($N_1 = 4900$, $N_2 = 280$).

\begin{table}[H]
    \centering
    \begin{tabular}{cccccc}
    \toprule
    Method&$m$ (DoFs)&$\Vert \mathbf{u} - \mathbf{u}_{\rho} \Vert_1$&$\Vert \mathbf{u} - \mathbf{u}_{\rho} \Vert_0$&$\Vert p - p_{\rho} \Vert_0$&$\Vert \nabla \cdot \mathbf{u} \Vert_0$ \\
    \midrule
    \multirow{5}{*}{RaNN}&$100\ (300)$&2.09E-03	&1.50E-04	&2.57E-03	&1.83E-03\\
    &$200\ (600)$&2.70E-05	&1.65E-06	&3.51E-05	&2.13E-05\\
    &$400\ (1200)$&3.91E-06	&2.25E-07	&4.81E-06	&3.10E-06\\
    &$800\ (2400)$&1.27E-06	&7.47E-08	&1.56E-06	&9.63E-07\\
    \midrule
    \multirow{5}{*}{SP-RaNN}&$100\ (200)$&2.09E-05	&9.40E-07	&2.35E-05	&1.46E-12\\
    &$200\ (400)$&8.02E-07	&3.00E-08	&1.00E-06	&5.73E-13\\
    &$400\ (800)$&1.79E-07	&6.32E-09	&1.87E-07	&2.63E-13\\
    &$800\ (1600)$&7.39E-08	&3.07E-09	&7.09E-08	&1.76E-13\\
    \bottomrule
    \end{tabular}
    \caption{Numerical errors for RaNN and SP-RaNN with different $m$, where $N_c=5180,\ r_1=1,\ r_2=0.5,\ \gamma=500$, for Re$=1$ in Example \ref{ex_2d_ns}).}
    \label{ex-stokes-re1}
\end{table}

\begin{table}[H]
\centering
\begin{tabular}{cccccc}
\toprule
Method&$m$ (DoFs)&$\Vert \mathbf{u} - \mathbf{u}_{\rho} \Vert_1$&$\Vert \mathbf{u} - \mathbf{u}_{\rho} \Vert_0$&$\Vert p - p_{\rho} \Vert_0$&$\Vert \nabla \cdot \mathbf{u} \Vert_0$ \\
\midrule
\multirow{5}{*}{RaNN}&$100\ (300)$&1.61E-03	&1.20E-04	&1.03E-06	&1.29E-05\\
&$200\ (600)$&4.16E-05	&2.34E-06	&2.28E-08	&5.14E-07\\
&$400\ (1200)$&1.24E-05	&6.43E-07	&6.01E-09	&1.94E-07\\
&$800\ (2400)$&5.86E-06	&2.86E-07	&3.04E-09	&1.14E-07\\
\midrule
\multirow{5}{*}{SP-RaNN}&$100\ (200)$&4.24E-06	&1.80E-07	&3.43E-09	&1.73E-12\\
&$200\ (400)$&2.29E-07	&9.36E-09	&2.68E-10	&5.84E-13\\
&$400\ (800)$&8.13E-08	&3.91E-09	&1.22E-10	&3.06E-13\\
&$800\ (1600)$&4.21E-08	&2.46E-09	&7.48E-11	&2.11E-13\\
\bottomrule
\end{tabular}
\caption{Numerical errors for RaNN and SP-RaNN with different $m$, where $N_c=5180,\ r_1=1,\ r_2=0.5,\ \gamma=1$, for Re$=1000$ in Example \ref{ex_2d_ns}).}
\label{ex-stokes-re1000}
\end{table}

\begin{figure}[!htbp] 		
	\centering
	\includegraphics[scale=0.75]{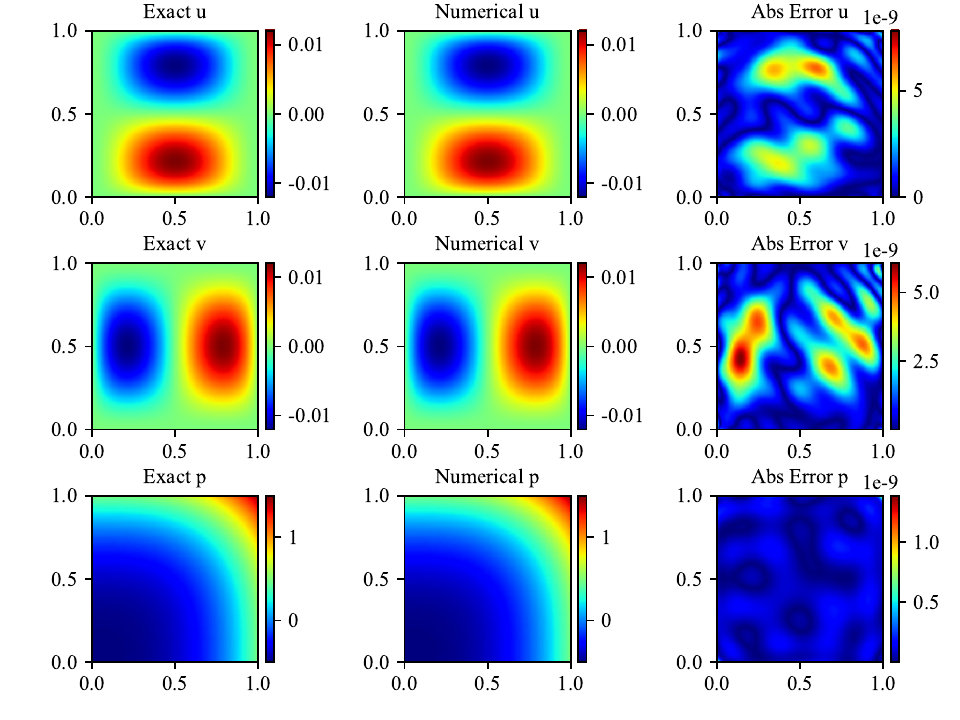}
	\caption{Exact solution, approximation solution of SP-RaNN, and absolute errors of $\mathbf{u}=(u,v)^T$, and $p$, where $m=800,\ \gamma=1,\ N_c=5180,\ r_1=1,\ r_2=0.5$ for Re$=1000$ in Example \ref{ex_2d_ns}.}
	\label{pp-stokes}
\end{figure}

The results for different Reynolds numbers are given in Tables \ref{ex-stokes-re1} and \ref{ex-stokes-re1000}. Figure \ref{pp-stokes} illustrates the exact solution, numerical solution, and absolute error. These results demonstrate that SP-RaNN achieves highly accurate numerical solutions. Several key observations are worth noting: SP-RaNN consistently achieves more accurate results with fewer DoFs compared to RaNN, as it satisfies the divergence-free constraint exactly. Moreover, as seen from Tables \ref{ex-stokes-re1} and \ref{ex-stokes-re1000}, SP-RaNN performs equally well for both high and low Reynolds numbers, indicating its robustness with respect to Reynolds number.

We also compare the proposed method with traditional numerical methods. In \cite{Linke2014}, a pressure-robust FEM method was proposed by modifying the test functions, ensuring that the velocity approximation error is independent of the pressure approximation. This method improves the accuracy of $\mathbf{u}$, especially at higher Reynolds numbers. On unstructured isotropic Delaunay meshes, the numerical results from \cite{Linke2014} are displayed in Table \ref{ex-stokes-fem}, extracted from Tables 4 and 6 in \cite{Linke2014}. It is evident that the accuracy and efficiency of the proposed SP-RaNN method surpass those of the FEM method in \cite{Linke2014}.

\begin{table}[H]
    \centering
    \begin{tabular}{ccccc}
    \toprule
    Re&DoFs&$| \mathbf{u}-\mathbf{u}_h |_{1}$&$\Vert \mathbf{u}-\mathbf{u}_h \Vert_0$&$\Vert p-p_h \Vert_0$ \\
    \midrule
    1& 102414&	1.98E-03&	5.91E-06&	4.0E-03 \\
    1000& 102414&	1.98E-03&	5.91E-06&	3.98E-03 \\
    \bottomrule
    \end{tabular}
    \caption{Numerical results from the mixed FEM with the modified Crouzeix–Raviart element (\cite{Linke2014}) in Example \ref{ex_2d_ns}.}
    \label{ex-stokes-fem}
\end{table}

\begin{example}[Unsteady Three-dimensional Beltrami Flow]\label{ex_3d_ns}
In this example, we solve the unsteady three-dimensional N-S equations \eqref{N-S-model}, following the settings as Example 3.3 in \cite{George2021}, where $\Omega = (-1, 1)^3$ and $I = (0, 1)$. The exact solution is as follows:
\begin{equation}
\left\{
\begin{array}{rrll}
    u &=& -(e^x {\rm sin}(y+z) + e^z {\rm cos}(x+y))e^{-t}, \\
    v &=& -(e^y {\rm sin}(z+x) + e^x {\rm cos}(y+z))e^{-t}, \\
    w &=& -(e^z {\rm sin}(x+y) + e^y {\rm cos}(z+x))e^{-t}, \\
    p &=& -\frac{1}{2}e^{-2t}[e^{2x} + e^{2y} + e^{2z} + 2{\rm sin}(x+y){\rm cos}(z+x)e^{y+z} \\
      &\ & + 2{\rm sin}(y+z){\rm cos}(x+y)e^{z+x} + 2{\rm sin}(z+x){\rm cos}(y+z)e^{x+y}]. \nonumber
\end{array}
\right.
\end{equation}
\end{example}

We use the SP-RaNN combined with Newton iteration to solve this problem. A SP-RaNN is used to approximate $\mathbf{u}$, and a separate RaNN approximates $p$, with parameters $r_1 = r_2 = 0.2$, $m = 3200$, $N_c = 6144$ ($N_1 = 4800$, $N_2 = 1344$), $\gamma = 100$, and ${\rm Re} = 1$. For this unsteady problem, 3 nonlinear iteration steps are used. Since the problem is time-dependent, we apply a constraint on $p$ at multiple time points, as explained in Remark \ref{constrain_p}. These points are randomly and uniformly selected from $I$, with 200 time points in total.

\begin{table}[H]
\centering
\begin{tabular}{ccccccc}
\toprule
Method&$e$&$t=0$&$t=0.25$&$t=0.5$&$t=0.75$&$t=1$ \\
\midrule
\multirow{5}{*}{SP-RaNN}&$e_{u_1}$&7.39E-06	&1.65E-06	&1.38E-06	&2.00E-06	&1.31E-05\\
&$e_{u_2}$&7.01E-06	&1.61E-06	&1.37E-06	&1.93E-06	&1.21E-05\\
&$e_{u_3}$&6.08E-06	&1.48E-06	&1.31E-06	&2.15E-06	&1.17E-05\\
&$e_{p}$&4.83E-05	&1.48E-05	&2.21E-05	&3.92E-05	&2.25E-04\\
&$\Vert \nabla \cdot \mathbf{u}\Vert_0$&2.05E-11	&2.08E-11	&2.09E-11	&2.11E-11	&2.09E-11\\
\midrule 
\multirow{4}{*}{NSFnets-VP}&$e_{u_1}$&1.10E-04&4.08E-04&5.94E-04&9.08E-04&1.63E-03 \\
&$e_{u_2}$&1.19E-04&5.48E-04&9.17E-04&1.31E-03&2.19E-03 \\
&$e_{u_3}$&1.14E-04&6.36E-04&9.87E-04&1.32E-03&1.78E-03 \\
&$e_{p}$&9.43E-03&8.98E-03&8.66E-03&8.56E-03&8.93E-03 \\
\midrule 
\multirow{3}{*}{NSFnets-VV}&$e_{u_1}$&7.40E-05&1.67E-04&2.23E-04&2.94E-04&4.35E-04 \\
&$e_{u_2}$&7.10E-05&1.53E-04&2.10E-04&3.02E-04&4.52E-04 \\
&$e_{u_3}$&7.40E-04&1.85E-04&2.43E-04&2.91E-03&4.09E-04 \\
\bottomrule
\end{tabular}
\caption{Comparison between NSFnets (Table 4 in \cite{George2021}) and SP-RaNN, where $m=3200,\ N_c=6144,\ r_1=r_2=0.2,\ \gamma=100$, for Re$=1$ in Example \ref{ex_3d_ns}.}
\label{compare-n-s-1}
\end{table}

\begin{table}[H]
\centering
\begin{tabular}{ccccccc}
\toprule
Re&$m$ (DoFs)&$e^{st}_{p}$&$e^{st}_{u}$&$e^{st}_{v}$&$e^{st}_{w}$\\
\midrule
\multirow{5}{*}{1}&100 (200)&2.91E-01	&1.17E-01	&1.08E-01	&1.33E-01\\
&200 (400)&1.61E-01	    &3.67E-02	&3.84E-02	&3.99E-02\\
&400 (800)&2.37E-02	    &5.50E-03	&5.68E-03	&5.41E-03\\
&800 (1600)&2.67E-03	&4.48E-04	&4.47E-04	&4.40E-04\\
&1600 (3200)&1.49E-04	&2.16E-05	&2.18E-05	&2.00E-05\\
\midrule
\multirow{5}{*}{1000}&100 (200)&1.91E-01	&1.10E-01	&1.17E-01	&1.18E-01\\
&200 (400)&5.58E-02	&3.31E-02	&3.16E-02	&3.53E-02\\
&400 (800)&1.00E-02	&5.22E-03	&5.22E-03	&4.96E-03\\
&800 (1600)&9.62E-04	&4.49E-04	&4.66E-04	&4.56E-04\\
&1600 (3200)&7.22E-05	&2.68E-05	&2.60E-05	&2.60E-05\\
\bottomrule
\end{tabular}
\caption{Space-time relative $L^2$ errors for SP-RaNN with different $m$, where $\ N_c=6144,\ r_1=r_2=0.2,\ \gamma=100$, for Re$=1$, and $\ N_c=6144,\ r_1=r_2=0.2,\ \gamma=10$, for $\text{Re} = 1000$ in Example \ref{ex_3d_ns}.}
\label{n-s-re1000}
\end{table}

\begin{figure}[!htbp] 		
	\centering
	\includegraphics[scale=0.75]{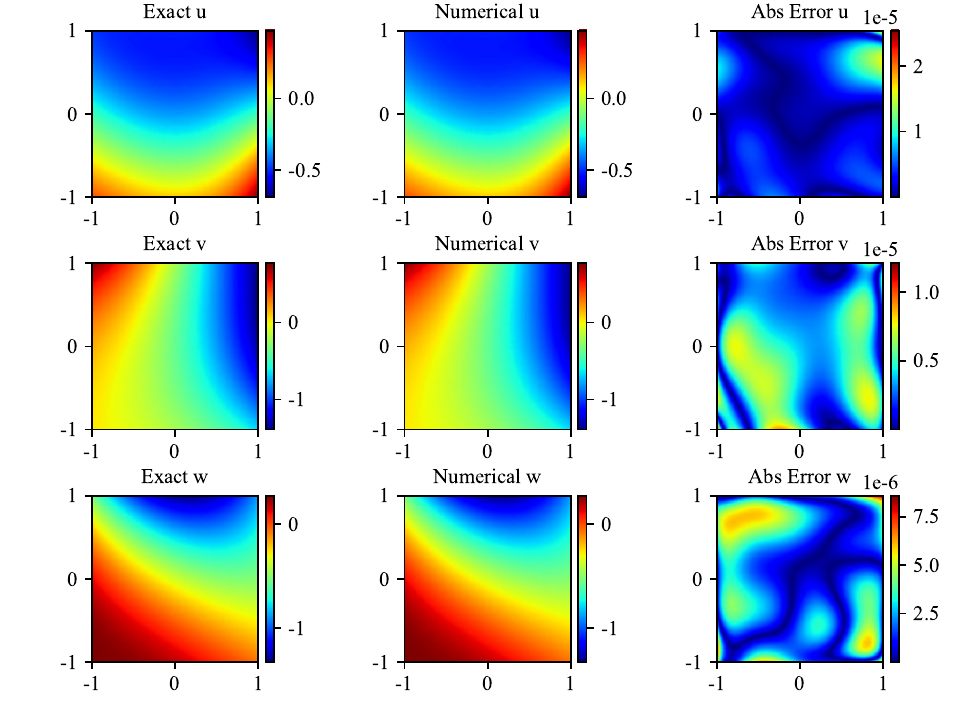}
	\caption{Exact solutions, approximation solutions and absolute errors of $\mathbf{u}=(u_1,u_2,u_3)^T$ at $t=1$ on the plane $z=0$, where $m=3200,\ \gamma=100,\ N_c=6144,\ r_1=r_2=0.2$, for ${\rm Re}=1$ in Example \ref{ex_3d_ns}.}
	\label{pp-n-s}
\end{figure}

We compare the performance of SP-RaNN against NSFnets (\cite{George2021}) in Table \ref{compare-n-s-1}, and display snapshots of $\mathbf{u}$ along with the absolute errors at $t = 1$ on the plane $z = 0$ in Figure \ref{pp-n-s}. The parameters and results for NSFnets are available in Table 4 of \cite{George2021}, where NN size is $7\times 50$, the number of collocation points is 15766. From Table \ref{compare-n-s-1}, it is evident that SP-RaNN achieves more accurate approximations, and the least-squares method enhances the effectiveness. Table \ref{n-s-re1000} shows significant improvements in accuracy as the number of neurons increases for SP-RaNN. The results for $\text{Re} = 1000$ further emphasize the robustness of the SP-RaNN method. These results validate the efficiency and robustness of SP-RaNN.

\begin{figure*}[!htbp] 		
	\centering    
	\subfigure[Physical setting]{\includegraphics[width=0.45\textwidth]{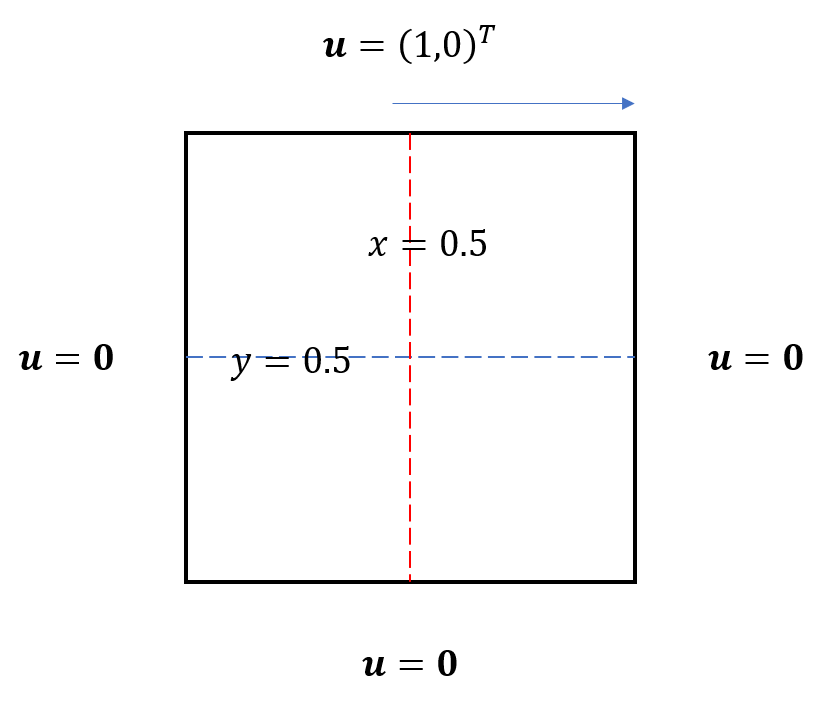}}
    \subfigure[Mexican hat wavelet]{\includegraphics[width=0.45\textwidth]{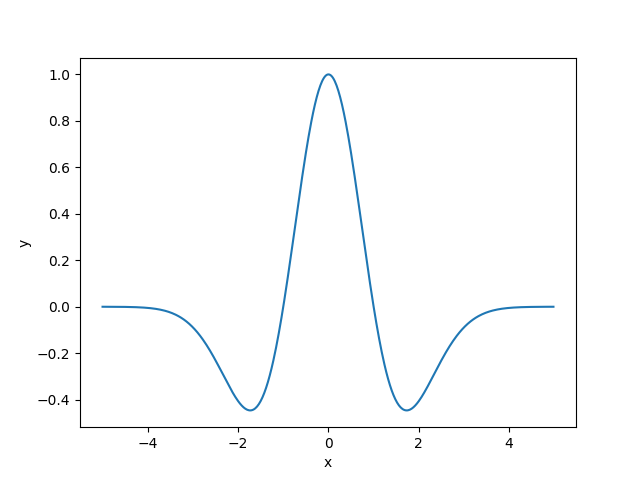}}
	\caption{The physical setting of lid-driven cavity flow and activation function in Example \ref{ex_ben_ns}.}
	\label{lid-driven-wavelet}
\end{figure*}

\begin{example}[Lid-driven Cavity Flow]\label{ex_ben_ns}
In this example, we test the proposed method on the lid-driven cavity flow problem:
\begin{equation} \label{ex_benchamrk}
    \left\{
    \begin{array}{rrll}
        (\mathbf{u}\cdot \nabla)\mathbf{u} + \nabla p - \frac{1}{{\rm Re}}\Delta \mathbf{u} &=& \mathbf{f},\quad& {\rm in}\ \Omega, \\
        \nabla \cdot \mathbf{u} &=& 0,\quad& {\rm in}\ \Omega. \\
    \end{array}
    \right.
\end{equation}
The domain is $\Omega = (0,1)^2$, with boundary conditions $\mathbf{u}(x, 1) = (1,0)^T$, $\mathbf{u}(x, 0) = \mathbf{u}(0, y) = \mathbf{u}(1, y) = \mathbf{0}$. To avoid discontinuities at the upper left and right corners, we set $\mathbf{u}(x, 1) = \big(1 - \frac{\cosh(50(x-0.5))}{\cosh(25)}, 0\big)^T$. The horizontal velocity $u$ on the vertical centerline (red dotted line) and vertical velocity $v$ on the horizontal centerline (blue dotted line) are test targets. The setup is illustrated in Figure \ref{lid-driven-wavelet} (a). 
\end{example}

\begin{figure*}[!htb]
    \centering
    \subfigure[$\rm Re=100$]{\includegraphics[width=0.329\textwidth]{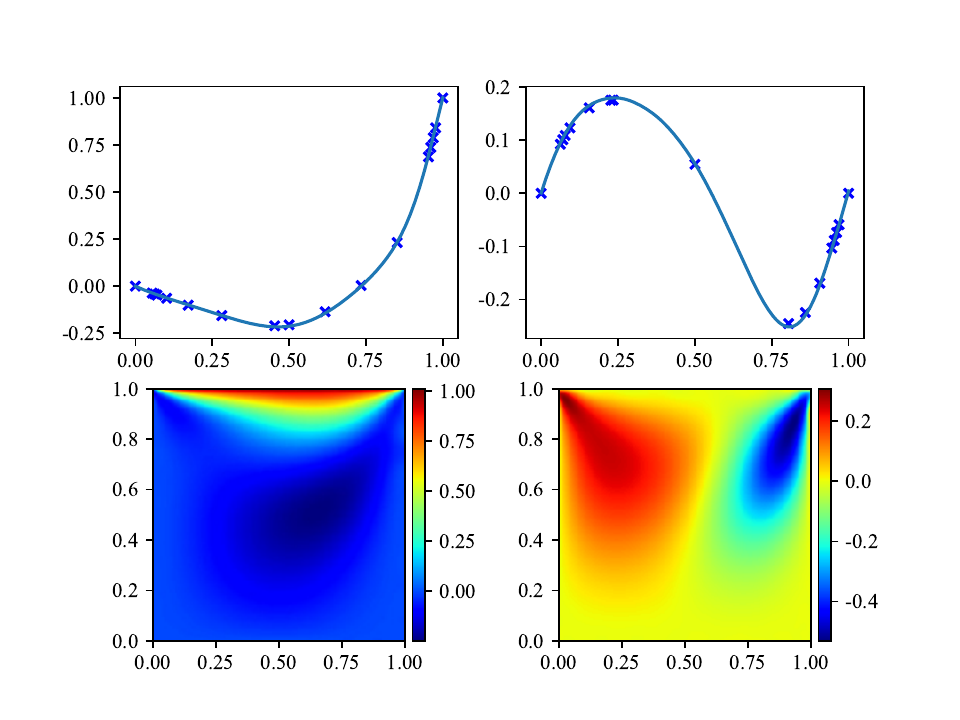}}
    \subfigure[$\rm Re=400$]{\includegraphics[width=0.329\textwidth]{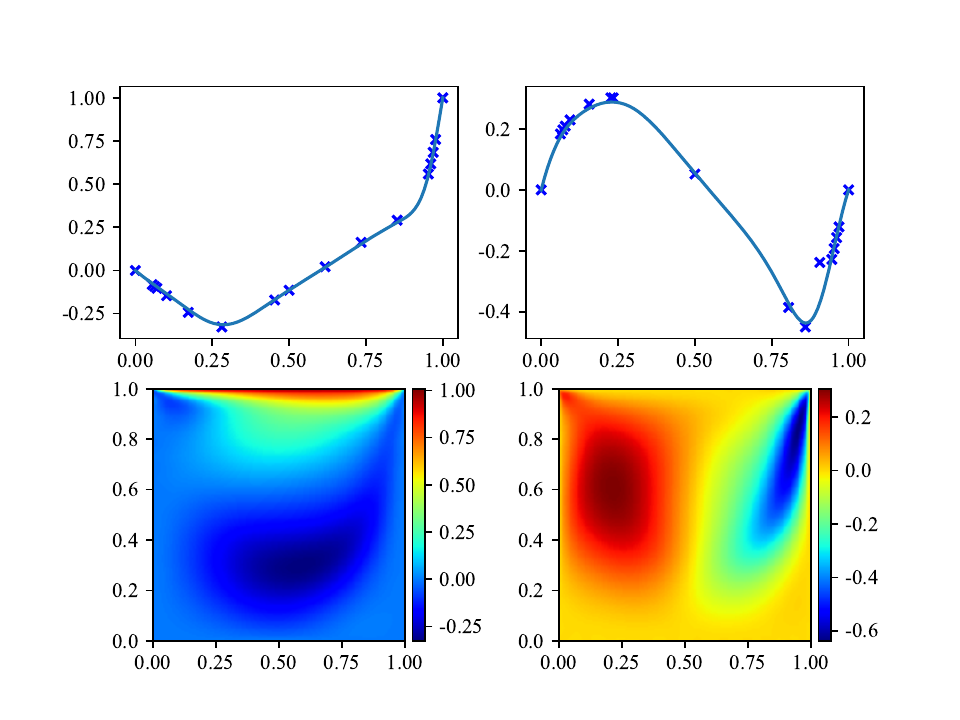}}
    \subfigure[$\rm Re=1000$]{\includegraphics[width=0.329\textwidth]{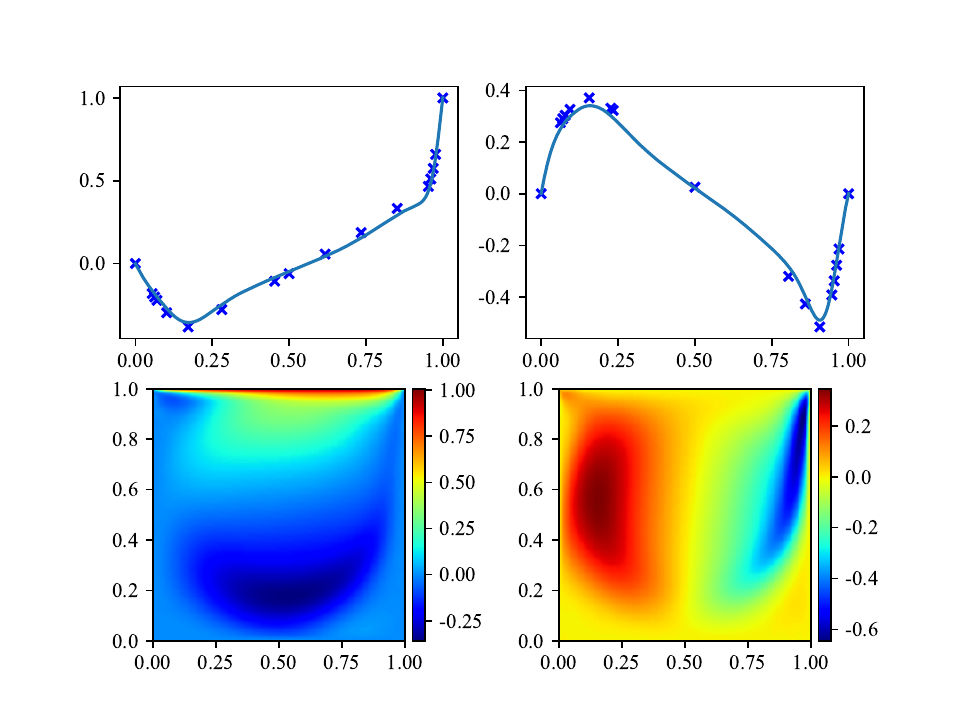}}
    \caption{Numerical solution $u$ and its values on vertical centerline, numerical solution $v$ and its values on horizontal centerline, where $r_1=6,\ r_2=3$ for Re$=100$, $r_1=10,\ r_2=5$ for Re$=400$ and $r_1=15,\ r_2=10$ for Re$=1000$ in Example \ref{ex_ben_ns}.}
    \label{benchmark_lid_driven}
    \vspace{0.1in}
\end{figure*}

\begin{figure*}[!htb]
    \centering
        \subfigure[Re=100]{\includegraphics[width=0.31\textwidth]{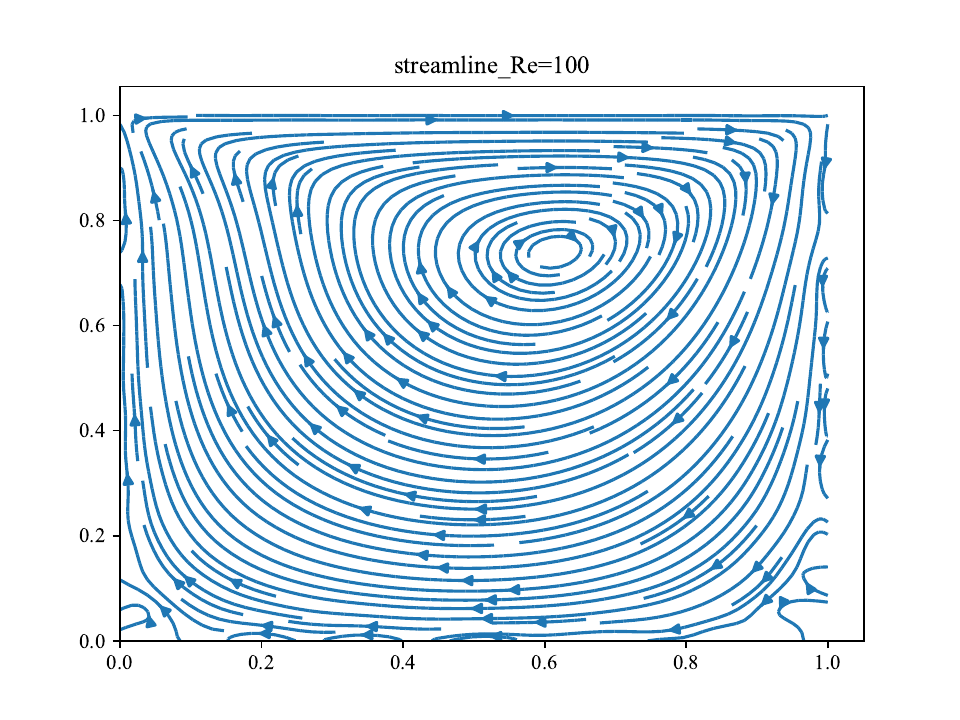}}
        \subfigure[Re=400]{\includegraphics[width=0.31\textwidth]{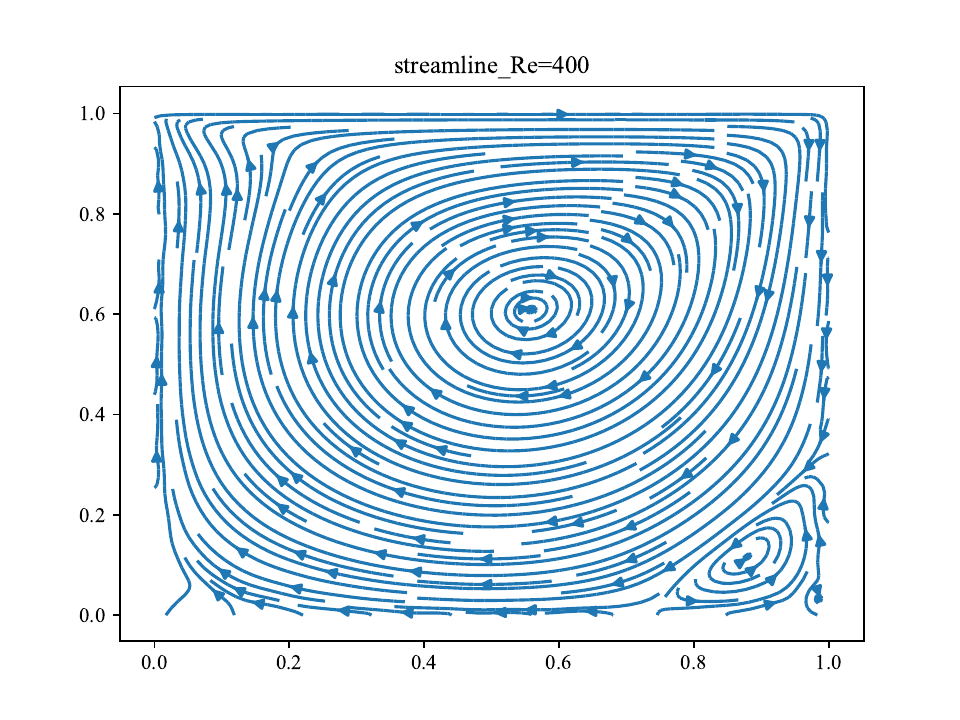}}
        \subfigure[Re=1000]{\includegraphics[width=0.31\textwidth]{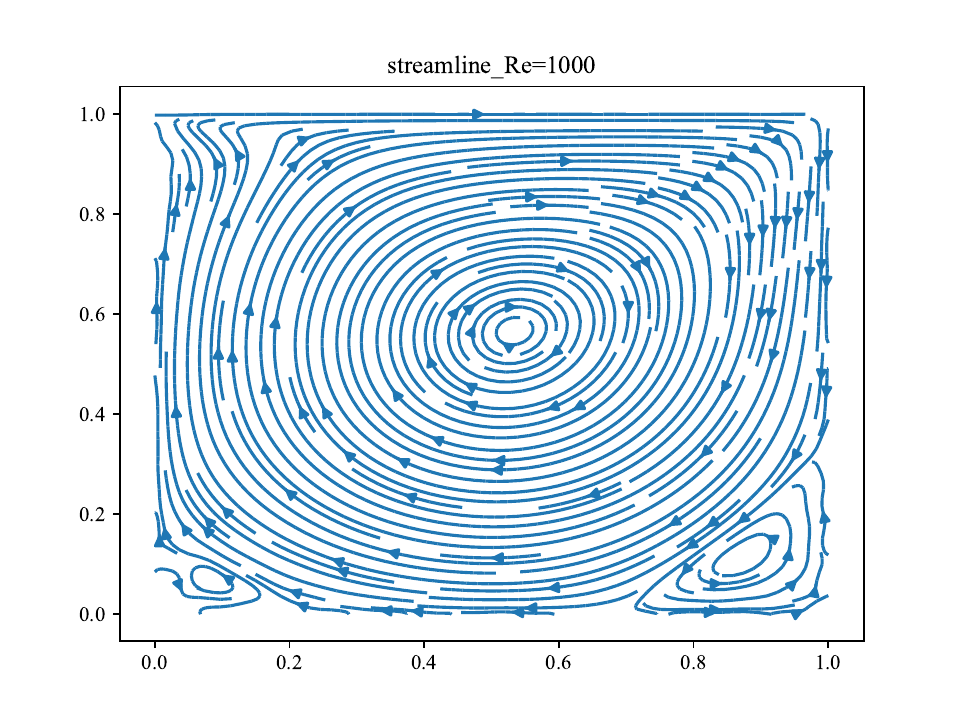}}\\
        \caption{Streamlines of the numerical solutions for different Reynolds numbers in Example \ref{ex_ben_ns}.}
        \label{streamline}
    \vspace{0.1in}
\end{figure*}

In this example, we use the Mexican hat wavelet $\rho = (1-x^2)e^{-\frac{x^2}{2}}$ as the activation function. This localized function (Figure \ref{lid-driven-wavelet} (b)) helps SP-RaNN capture the flow characteristics more effectively. A SP-RaNN with parameter $r_1$ is used to approximate the velocity $\mathbf{u}$, and an RaNN with parameter $r_2$ approximates the pressure $p$. The parameters are set to $m = 1000$, $N_c = 5180$, $\gamma = 10$, and Newton iteration is used. The results for different Reynolds numbers are shown in Figure \ref{benchmark_lid_driven}. The left-hand plots show horizontal velocities $u$, and the right-hand plots show vertical velocities $v$, with simulated results (blue curves) compared to reference values (marked with ``$\times$"). Streamlines for different Reynolds numbers are shown in Figure \ref{streamline}, and Tables 1 and 2 from \cite{Ghia1982} serve as the reference.

For $\text{Re} = 100, 400, 1000$, the number of iteration steps is 10, 15, and 25, respectively. Higher Reynolds numbers increase the nonlinearity of the equation, requiring more iterations. We employ the initialization strategy from \cite{Dang2024} to determine the bias. This method concentrates more basis functions within $\Omega$, using a weight matrix $W \in \mathbb{R}^{m \times 2}$, and generating a random matrix $B \in \mathbb{R}^{m \times 2}$. The bias $\mathbf{b}$ is then selected as $\mathbf{b} = -(W \odot B) \cdot \mathbf{1}_{2 \times 1}$, where $\odot$ denotes element-wise multiplication.

\begin{example}[Unsteady Three-Dimensional Maxwell Equations]\label{ex_maxwell}

In this example, we utilize the SP-RaNN to solve the unsteady three-dimensional Maxwell equations \eqref{maxwell-model}. The domain is set as $\Omega=(0,1)^3,\ I=(0,1)$, with $\epsilon=\mu=1$. The exact solution is provided as follows:
\begin{equation}
\left\{
\begin{array}{rrll}
    E_1 &=& (y-y^2)(z-z^2)t{\rm cos}(t+x+y+z),  \\
    E_2 &=& (x-x^2)(z-z^2)t{\rm cos}(t+x+y+z), \\
    E_3 &=& (x-x^2)(y-y^2)t{\rm cos}(t+x+y+z), \nonumber
\end{array}
\right.
\end{equation}
and
\begin{equation}
\left\{
\begin{array}{rrll}
    H_1 &=& (y-z)t{\rm cos}(t+x+y+z),  \\
    H_2 &=& (z-x)t{\rm cos}(t+x+y+z), \\
    H_3 &=& (x-y)t{\rm cos}(t+x+y+z). \nonumber
\end{array}
\right.
\end{equation}
\end{example}

\begin{figure}[!htbp] 		
	\centering
	\includegraphics[scale=0.75]{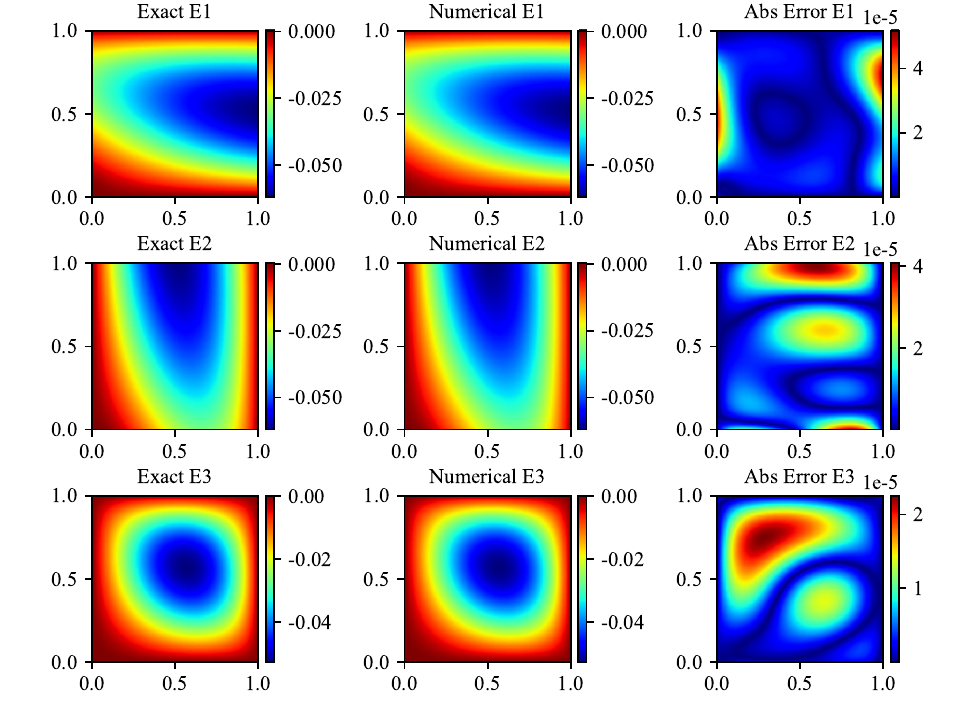}
	\caption{Electric fields and absolute error at $t=1$ on the plane $z=0.5$, where $\ m=1600,\ \gamma=100,\ N_c=6144,\ r_1=0.5$ and $ r_2=0.2$ in Example \ref{ex_maxwell}.}
	\label{pp-maxwell-E}
\end{figure}

\begin{figure}[!htbp] 		
	\centering
	\includegraphics[scale=0.75]{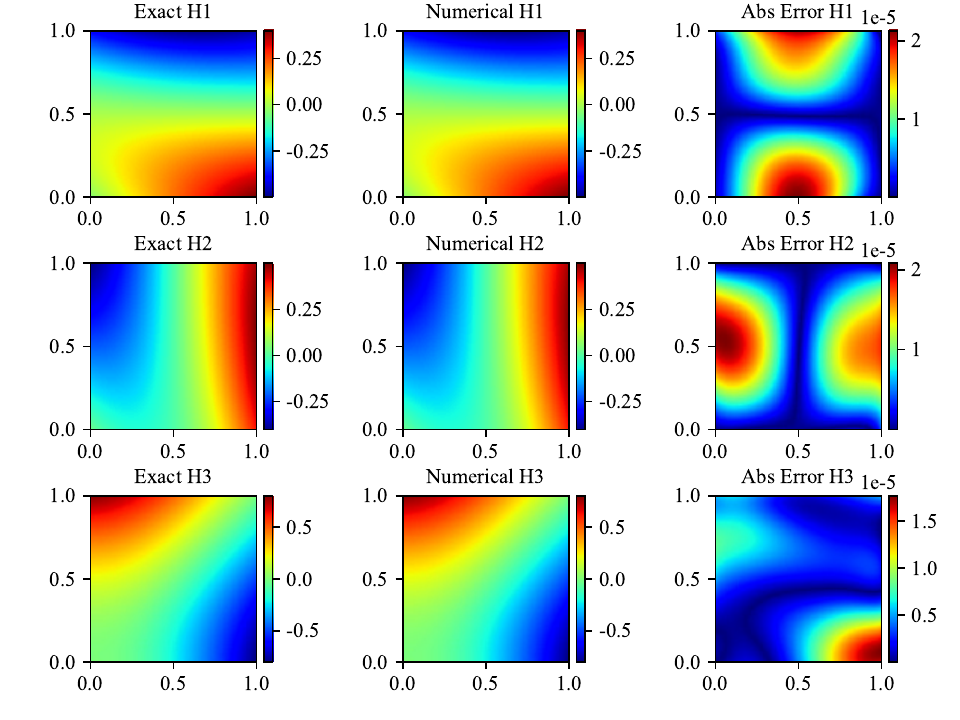}
	\caption{Magnetic fields and absolute error at $t=1$ on the plane $z=0.5$, where $\ m=1600,\ \gamma=100,\ N_c=6144,\ r_1=0.5$ and $ r_2=0.2$ in Example \ref{ex_maxwell}.}
	\label{pp-maxwell-H}
\end{figure}

This example was previously addressed using the space-time discontinuous Galerkin (ST-DG) method with the uniform Cartesian mesh in \cite{Xie2013} (Example 4.2). In our approach, a RaNN with $r_1 = 0.5$ is used to approximate $\mathbf{E}$, and a SP-RaNN with $r_2 = 0.2$ is employed to approximate $\mathbf{H}$. The parameters are set to $N_c = 6144$ ($N_1 = 4800$, $N_2 = 1344$) and $\gamma = 100$. Table \ref{RaNN-maxwell} shows the $L^2$ errors for $\mathbf{E}$ and $\mathbf{H}$ across degrees of freedom, comparing the SP-RaNN and ST-DG methods with different polynomial orders for time and space discretization ($r=k=1$ and $r=k=2$). The results demonstrate that SP-RaNN achieves more accurate approximations with fewer DoFs while maintaining the divergence-free condition. Figures \ref{pp-maxwell-E} and \ref{pp-maxwell-H} display the exact solutions, numerical approximations, and absolute errors.

\begin{table}[H]
\centering
\begin{tabular}{ccccc}
\toprule
Methods&DoFs&$\Vert \mathbf{E}-\mathbf{E}_{\rho}\Vert_0$&$\Vert \mathbf{H}-\mathbf{H}_{\rho}\Vert_0$&$\Vert \nabla \cdot \mathbf{H}_{\rho}\Vert_0$\\
\midrule
\multirow{5}{*}{SP-RaNN}&400&1.76E-02	&6.54E-02	&4.23E-14\\
&800&1.05E-02	&2.23E-02	&1.97E-13\\
&1600&2.15E-03	&2.19E-03	&1.00E-12\\
&3200&3.78E-04	&2.35E-04	&6.42E-12\\
&6400&5.27E-05	&1.26E-05	&7.46E-12\\
\midrule
& &$\Vert \mathbf{E}-\mathbf{E}_{h}\Vert_0$&$\Vert \mathbf{H}-\mathbf{H}_{h}\Vert_0$&$\Vert \nabla \cdot \mathbf{H}_h\Vert_0$\\
\midrule
\multirow{2}{*}{ST-DG-$P_1$}&147456&1.16E-02	&1.44E-02	&-\\
&2359296&2.95E-03	&3.90E-03	&-\\
\midrule
\multirow{2}{*}{ST-DG-$P_2$}&746496&9.89E-04	&1.06E-03	&-\\
&11943936&1.23E-04	&1.38E-04	&-\\
\bottomrule
\end{tabular}
\caption{$L^2$ errors at $t=1$, where $N_c=6144,\,r_1=0.5,\ r_2=0.2,\ \gamma=100$ in Example \ref{ex_maxwell}.}
\label{RaNN-maxwell}
\end{table}

\begin{example}[Three-Dimensional Simplified Maxwell and MHD Equations]\label{ex_time_test}
In this example, we compare the computational efficiency of the proposed method with that of a preconditioned FEM for the simplified Maxwell equation:
\begin{equation} \label{ex-maxwell-s}
\left\{
\begin{array}{rrll}
    \nabla \times \nabla \times \mathbf{A} + \mathbf{A} &=& \mathbf{f},\ & {\rm in}\ \Omega,\\
    \mathbf{A}\times \mathbf{n} &=& \mathbf{g}\times \mathbf{n},\ & {\rm on}\ \partial \Omega,\\
\end{array}
\right.
\end{equation}
and the simplified magnetohydrodynamics equations
\begin{equation} \label{ex-MHD-s}
\left\{
\begin{array}{rrll}
    \nabla p - \frac{1}{{\rm Re}}\Delta \mathbf{u} - {\rm S}(\nabla \times \mathbf{B})\times\mathbf{B} &=& \mathbf{f},\quad& {\rm in}\ \Omega, \\
    \frac{1}{\rm R_m}\nabla \times(\nabla \times \mathbf{B}) - \nabla \times(\mathbf{u} \times \mathbf{B}) &=& \mathbf{g},\quad& {\rm in}\ \Omega, \\
    \nabla \cdot \mathbf{u}\;\;\; =\;\;\; 0,\quad  \nabla \cdot \mathbf{B} &=& 0, & {\rm in}\ \Omega, \\
    \mathbf{u} \;\;\;=\;\;\; \mathbf{u}_{D},\quad  \mathbf{B} \times \mathbf{n} &=& \mathbf{B}_{D}, & {\rm on}\ \partial \Omega, \\
\end{array}
\right.
\end{equation}
where $\Omega=(0,1)^3$, ${\rm Re}={\rm R_m}={\rm S}=1$.

\end{example}

\begin{itemize}
    \item Simplified Maxwell Equations \eqref{ex-maxwell-s}
\end{itemize}
The exact solution is 
$$\mathbf{A}=\big((y^2+z^2){\rm sin}(\pi x),\ (x^2+z^2){\rm sin}(\pi y),\ (x^2+y^2){\rm sin}(\pi z)\big)^T.$$ 
The reference FEM computation uses tetrahedral meshes with first-order N\'ed\'elec edge elements (Type~II) for discretizing $\mathbf{A}$.  
The resulting linear system is solved by the Preconditioned Conjugate Gradient method combined with the Hiptmair--Xu auxiliary-space preconditioner (\cite{Hiptmair2007}), which is widely regarded as an optimal solver for $H(\mathrm{curl})$-conforming problems.

For comparison, a single RaNN is employed to approximate $\mathbf{A}$, with parameters $r=1$, $N_c = 4725$, and $\gamma = 100$.  
Despite the FEM using advanced preconditioning, RaNN achieves comparable or higher accuracy at a fraction of the computational cost, as shown in Table~\ref{time-test-maxwell}.

\begin{table}[H]
    \centering
    \begin{tabular}{cccc}
    \toprule
    Method&DoFs&$\Vert \mathbf{A} - \mathbf{A}_{\rho} \Vert_0$&Running time (s)\\
    \midrule
    \multirow{4}{*}{FEM}&8368&1.12E-02	&0.12\\
    &62048&2.80E-03	&1.02\\
    &477376&6.98E-04	&8.51\\
    &3744128&1.74E-04	&76.53\\
    \midrule
    \multirow{4}{*}{RaNN}&300&2.76E-02	&0.14\\
    &600&1.90E-03	&0.29\\
    &1200&2.45E-05	&0.78\\
    &2400&1.94E-06	&2.44\\
    \bottomrule
    \end{tabular}
        \caption{Comparison between the preconditioned FEM and RaNN for Equation~\eqref{ex-maxwell-s}. Numerical errors and running times are reported for different DoFs, where $N_c = 4725$, $r = 1$, and $\gamma = 100$.}
    \label{time-test-maxwell}
\end{table}

\begin{itemize}
    \item Simplified MHD Equations \eqref{ex-MHD-s}
\end{itemize}
The exact solution is 
$$\mathbf{u}=\big(\sin(z),\ 2\cos(x),\ 0\big)^T,\quad p=\sin(y)-1+\cos(1),\quad \mathbf{B}=\big(\cos(y),0,0\big)^T.$$ 
For the FEM baseline, we introduce a Lagrange multiplier $r$ to numerically enforce the divergence-free constraint on the magnetic field $\mathbf{B}$. Second-order Lagrange elements are employed for $\mathbf{u}$, first-order Lagrange elements for $p$, first-order N\'ed\'elec edge elements for $\mathbf{B}$, and second-order Lagrange elements for $r$.  
The resulting large-scale saddle-point system is solved using the Flexible GMRES algorithm with a block preconditioner following \cite{Li2017}, consistent with state-of-the-art MHD solvers.

In the proposed approach, two SP-RaNNs are employed to approximate $\mathbf{u}$ and $\mathbf{B}$, and one RaNN is used to approximate $p$, with parameters $r_1 = r_2 = r_3 = 1$, $N_c = 1600$, and $\gamma = 100$.  
Table~\ref{time-test-mhd} shows that the SP-RaNN method attains comparable accuracy while being orders of magnitude faster than the preconditioned FEM baseline, highlighting its efficiency.

\begin{table}[H]
    \centering
    \begin{tabular}{ccccccccc}
    \toprule
    Method&DoFs&$\Vert \mathbf{u} - \mathbf{u}_{\rho} \Vert_1$&$\Vert p - p_{\rho} \Vert_0$&$\Vert \mathbf{B} - \mathbf{B}_{\rho} \Vert_{\rm H(curl)}$&$\Vert \nabla \cdot \mathbf{u} \Vert_0$&$\Vert \nabla \cdot \mathbf{B} \Vert_0$& $N$ & Running time (s)\\
    \midrule
    \multirow{4}{*}{FEM}&4249&2.90E-03	&1.84E-03	&4.81E-02	&- &- &8 &1.72\\
    &28749&7.08E-04	&3.90E-04	&2.38E-02	&- &- &8 &15.12\\
    &210709&1.75E-04	&9.07E-05	&1.18E-02	&- &- &8 &135.46\\
    &1611813&4.34E-05	&2.21E-05	&5.89E-03	&- &- &7 &981.22\\
    \midrule
    \multirow{5}{*}{SP-RaNN}&300&1.39E-02	&1.43E-02	&6.55E-03	&5.22E-16 &2.10E-16 &5 &0.60\\
    &600&1.56E-03	&1.78E-03	&5.58E-04	&9.35E-16 &4.12E-16 &5 &1.39\\
    &1200&6.75E-05	&7.98E-05	&2.27E-05	&1.24E-16 &5.15E-16 &5 &2.96\\
    &2400&1.39E-06	&2.72E-06	&5.20E-07	&1.89E-16 &6.46E-16 &5 &9.46\\
    &4800&1.56E-08	&5.08E-08	&8.34E-09	&2.07E-15 &1.02E-15 &5 &63.83\\
    \bottomrule
    \end{tabular}
        \caption{Comparison between the preconditioned FEM and SP-RaNN for Equation~\eqref{ex-MHD-s}. Numerical errors and running times are reported for different DoFs, where $N_c = 1600$, $r_1 = r_2 = r_3 = 1$, and $\gamma = 100$. Here, $N$ denotes the number of Picard iterations.}
    \label{time-test-mhd}
\end{table}

\begin{example}[Steady Two-Dimensional Magnetohydrodynamics Equations]\label{ex_2d_mhd}

In this example, we compare our method with the robust globally divergence-free weak Galerkin finite element method (WG-FEM) from \cite{Xie2024}, which guarantees divergence-free velocity and magnetic field approximations. We consider the steady incompressible MHD equations \eqref{ex-mhd-2d}:
\begin{equation} \label{ex-mhd-2d}
\left\{
\begin{array}{rrll}
    -\frac{1}{\rm H_a^2}\Delta \mathbf{u} + \frac{1}{N}(\mathbf{u}\cdot \nabla) \mathbf{u} + \nabla p - \frac{1}{\rm R_m}(\nabla \times \mathbf{B})\times \mathbf{B} &=& \mathbf{f}, \\
    \frac{1}{\rm R_m} \nabla \times (\nabla \times \mathbf{B}) - \nabla \times (\mathbf{u} \times \mathbf{B}) &=& \mathbf{g}, \\
    \nabla \cdot \mathbf{u} &=& 0,\\
    \nabla \cdot \mathbf{B} &=& 0,
\end{array}
\right.
\end{equation}
where $\rm H_a,\ N,\ R_m$ are the Hartmann number, the interaction parameter and the magnetic Reynolds number, respectively. 
\end{example}
\begin{itemize}
\item Case 1: Simply connected domain
\end{itemize}
We follow the same test case as in \cite{Xie2024} (Example 7.1), where the domain is $\Omega=(0,1)^2$, and parameters $H_a=N=R_m=1$. The exact solution is:
\begin{equation}
\left\{
\begin{array}{l}
    u =  x^2 (x-1)^2 y (y-1) (2y-1),\ v = y^2 (y-1)^2 x (x-1) (2x-1),  \\
    B_1 =  x^2 (x-1)^2 y (y-1) (2y-1),\ B_2 = y^2 (y-1)^2 x (x-1) (2x-1), \\
    p = x (x-1) (x-\frac{1}{2}) y (y-1) (y-\frac{1}{2}). \nonumber
\end{array}
\right.
\end{equation}

\begin{figure}[!htbp] 		
	\centering
	\includegraphics[scale=0.75]{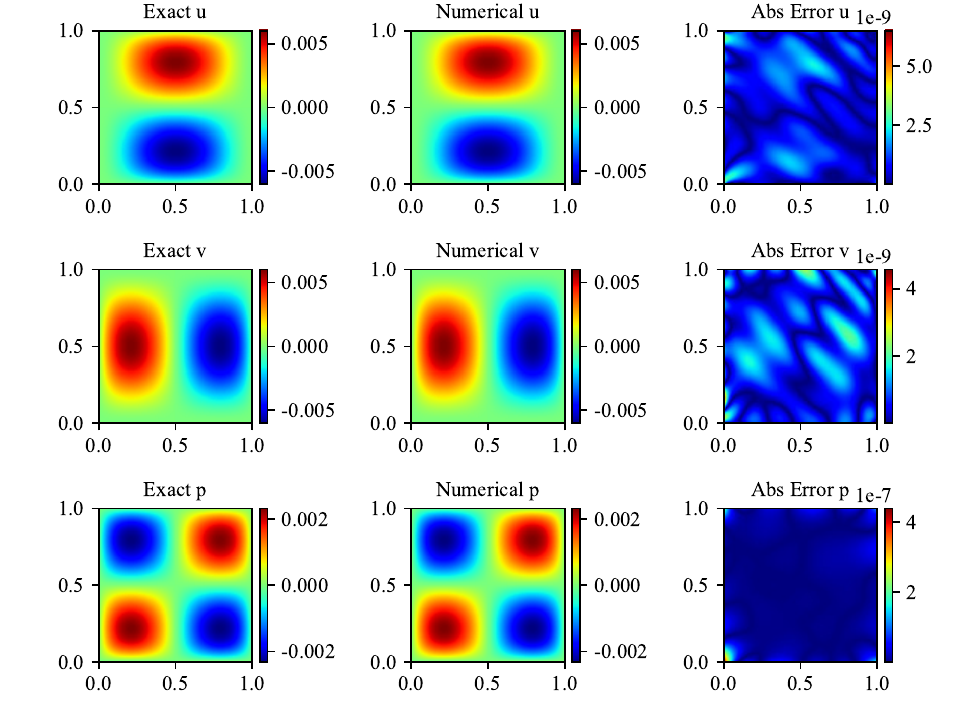}
	\caption{Velocity field, pressure field and absolute error, where $\ m=800,\ \gamma=100,\ N_c=5180,\ r_1=r_2=1$ and $r_3=0.5$ in Example \ref{ex_2d_mhd}, case 1.}
	\label{pp-mhd}
\end{figure}

We apply SP-RaNN with both Picard iteration (SP-RaNN-P) and Newton iteration (SP-RaNN-N) separately. Two SP-RaNNs with $r_1=1$ and $r_2=1$, and a RaNN with $r_3=0.5$, are used to approximate $\mathbf{u}$, $\mathbf{B}$, and $p$, respectively. The parameters are set as $N_c = 5180\ (N_1=4900,\ N_2=280)$ and $\gamma=100$. We apply 3-step iteration to address the nonlinear terms. Table \ref{compare-mhd-2d} shows the results obtained by WG-FEM with $k=1$ and $k=2$, on regular triangular meshes, and SP-RaNN using both nonlinear iteration methods. Both methods achieve nearly identical results while maintaining the exact divergence-free condition, achieving higher accuracy with fewer DoFs compared to WG-FEM. The exact solutions, simulations, and absolute errors of $\mathbf{u}$ and $p$ are shown in Figure \ref{pp-mhd}.

To further verify boundary robustness, we impose the normal-component magnetic boundary condition ${\bf B}\cdot {\bf n} = 0$ instead of the tangential condition ${\bf B}\times {\bf n}$, and evaluate the SP-RaNN-P method using the same network architecture and parameters ($r_1=r_2=1,\ r_3=0.5,\ N_c=5180$). Table \ref{change-gamma} reports results for several values of $\gamma$, demonstrating that the method effectively enforces ${\bf B\cdot n}=0$ and remains insensitive to the choice of $\gamma$. Increasing the boundary and initial condition weights generally improves accuracy, though excessively large $\gamma$ may cause overfitting and reduce overall performance.

\begin{table}[H]
\centering
\begin{tabular}{ccccccccc}
\toprule
$\gamma$ &DoFs&$e_{\mathbf{u}}$&$e_{\nabla \mathbf{u}}$&$\Vert \nabla \cdot \mathbf{u} \Vert_0$&$e_{\mathbf{B}}$&$e_{\nabla \mathbf{B}}$&$\Vert \nabla \cdot \mathbf{B} \Vert_0$&$e_{p}$ \\
\midrule
\multirow{5}{*}{$10$}&$300$&1.03E-03&	2.48E-03&	6.71E-13&	1.35E-02&	1.02E-02&	9.18E-13&	5.67E-02\\
&$600$&5.68E-05&	1.74E-04&	3.39E-13&	1.65E-03&	1.20E-03&	3.24E-13&	4.93E-03\\
&$1200$&8.75E-06&	2.95E-05&	1.16E-13&	3.65E-04&	2.66E-04&	1.28E-13&	7.83E-04\\
&$2400$&2.84E-06&	9.97E-06&	9.30E-14&	9.25E-05&	6.85E-05&	9.53E-14&	2.68E-04\\

\midrule
\multirow{5}{*}{$10^2$}&$300$&1.28E-04&	3.87E-04&	8.89E-13&	3.29E-03&	2.42E-03&	7.46E-13&	1.03E-02\\
&$600$&4.13E-06&	1.49E-05&	2.88E-13&	6.16E-05&	4.60E-05&	2.92E-13&	4.50E-04\\
&$1200$&8.40E-07&	3.11E-06&	1.56E-13&	3.22E-05&	2.43E-05&	1.61E-13&	7.34E-05\\
&$2400$&3.78E-07&	1.28E-06&	9.02E-14&	2.57E-05&	1.90E-05&	9.08E-14&	2.73E-05\\

\midrule

\multirow{5}{*}{$10^3$}&$300$&2.03E-05&	6.49E-05&	9.86E-13&	5.17E-04&	3.95E-04&	9.75E-13&	1.28E-03\\
&$600$&1.02E-06&	3.44E-06&	2.43E-13&	1.74E-05&	1.33E-05&	3.15E-13&	5.47E-05\\
&$1200$&3.77E-07&	1.10E-06&	1.39E-13&	7.27E-06&	5.49E-06&	1.62E-13&	1.78E-05\\
&$2400$&2.63E-07&	5.39E-07&	9.46E-14&	3.45E-06&	2.84E-06&	1.09E-13&	8.76E-06\\

\midrule
\multirow{5}{*}{$10^4$}&$300$&5.41E-05&	1.35E-04&	1.10E-12&	4.49E-04&	3.50E-04&	1.21E-12&	1.84E-03\\
&$600$&1.47E-06&	4.39E-06&	3.07E-13&	1.59E-05&	1.25E-05&	3.26E-13&	5.97E-05\\
&$1200$&4.18E-07&	1.16E-06&	1.40E-13&	6.63E-06&	5.36E-06&	1.58E-13&	1.87E-05\\
&$2400$&2.75E-07&	6.53E-07&	1.06E-13&	4.39E-06&	3.70E-06&	9.60E-14&	1.08E-05\\
\bottomrule
\end{tabular}
\caption{Numerical errors for various values of $\gamma$ with the boundary condition ${\bf B}\cdot {\bf n}=0$, where $N_c=5180,\ r_1=r_2=1,\ r_3=0.5$ in Example \ref{ex_2d_mhd}.}
\label{change-gamma}
\end{table}

Additionally, we test this example for various Hartmann numbers, using 10 Newton iterations with parameters $r_1 = r_2 = 1$, $r_3 = 0.5$, $N = {\rm R_m} = 1$. The results are presented in Table \ref{hartmann_change}.

\begin{table}[H]
\centering
\begin{tabular}{ccccccccc}
\toprule
Method&DoFs&$e_{\mathbf{u}}$&$e_{\nabla \mathbf{u}}$&$\Vert \nabla \cdot \mathbf{u} \Vert_0$&$e_{\mathbf{B}}$&$e_{\nabla \mathbf{B}}$&$\Vert \nabla \cdot \mathbf{B} \Vert_0$&$e_{p}$ \\
\midrule
\multirow{5}{*}{SP-RaNN-P}&$300$&9.92E-05	&3.10E-04	&8.36E-13	&7.70E-05	&2.45E-04	&8.63E-13	&8.81E-03\\
&$600$&3.53E-06	    &1.28E-05	&2.62E-13	&3.11E-06	&9.07E-06	&2.75E-13	&3.34E-04\\
&$1200$&8.31E-07	    &3.30E-06	&1.26E-13	&1.31E-06	&3.19E-06	&1.82E-13	&8.41E-05\\
&$2400$&3.52E-07	    &1.19E-06	&8.44E-14	&8.61E-07	&1.19E-06	&9.87E-14	&2.72E-05\\
\midrule
\multirow{5}{*}{SP-RaNN-N}&$300$&9.30E-05	&2.75E-04	&4.80E-13	&8.97E-05	&2.77E-04	&1.12E-12	&7.50E-03\\
&$600$&3.66E-06	    &1.26E-05	&2.59E-13	&4.11E-06	&9.87E-06	&2.75E-13	&3.63E-04\\
&$1200$&7.41E-07	    &2.69E-06	&1.58E-13	&1.19E-06	&2.83E-06	&1.71E-13	&5.69E-05\\
&$2400$&4.13E-07	    &1.34E-06	&9.13E-14	&8.74E-07	&1.32E-06	&1.18E-13	&3.05E-05\\
\midrule
\multirow{2}{*}{WG-FEM ($k=1$)}&$148992$&2.59E-03&4.02E-02&3.42E-14&4.63E-03&3.69E-02&2.09E-14&3.01E-02 \\
&$592896$&6.52E-04&2.01E-02&2.96E-13&1.18E-03&1.83E-02&2.98E-13&1.50E-02 \\
\midrule
\multirow{2}{*}{WG-FEM ($k=2$)}&$223448$&1.47E-05&9.65E-04&5.22E-14&7.82E-05&5.39E-03&6.05E-14&2.78E-04 \\
&$889344$&1.84E-06&2.41E-04&6.99E-13&9.90E-06&1.35E-03&6.79E-13&6.96E-05 \\
\bottomrule
\end{tabular}
\caption{Comparison of WG-FEM (Table 4 in \cite{Xie2024}) and SP-RaNN, where $N_c=5180,\ r_1=r_2=1,\ r_3=0.5$ and $\gamma=100$ in Example \ref{ex_2d_mhd}.}
\label{compare-mhd-2d}
\end{table}

\begin{table}[H]
    \centering
    \begin{tabular}{ccccccccc}
    \toprule
    $\rm H_a$&DoFs&$e_{\mathbf{u}}$&$e_{\nabla \mathbf{u}}$&$\Vert \nabla \cdot \mathbf{u} \Vert_0$&$e_{\mathbf{B}}$&$e_{\nabla \mathbf{B}}$&$\Vert \nabla \cdot \mathbf{B} \Vert_0$&$e_{p}$ \\
    \midrule
    10&$2400$&5.72E-07	&1.26E-06	&2.08E-13	&8.50E-07	&1.14E-06	&1.10E-13	&1.10E-06\\
    100&$2400$&5.08E-04	&8.91E-04	&4.49E-11	&8.94E-07	&1.37E-06	&9.19E-14	&8.39E-06\\
    1000&$2400$&4.81E-03	&7.89E-03	&4.95E-11	&1.05E-06	&1.38E-06	&1.01E-13	&8.42E-05\\
    \bottomrule
    \end{tabular}
    \caption{Errors for different Hartmann numbers using SP-RaNN-N, with $N_c=5180,\ r_1=r_2=1,\ r_3=0.5$ and $\gamma=100$ in Example \ref{ex_2d_mhd}.}
    \label{hartmann_change}
    \end{table}

\begin{itemize}
\item Case 2: Annular domain
\end{itemize}
We consider a non‑simply‑connected domain \(\Omega = \{\mathbf{x} \mid 0.5 < |\mathbf{x}| < 1\}\). The exact solution and parameters are the same as in Case 1, and the SP-RaNN-P method is used to solve this problem. As before, two SP-RaNNs with \(r_1 = r_2 = 1\) are employed for \(\mathbf{u}\) and \(\mathbf{B}\), and an RaNN with \(r_3 = 0.5\) is used for \(p\). We set \(N_c = 5460\) (\(N_1 = 4900, N_2 = 560\)), \(\gamma = 100\), and perform 3 iteration steps. The numerical results are summarized in Table \ref{mhd-2d-non}.  

\begin{figure}[!htbp] 		
	\centering
	\includegraphics[scale=0.75]{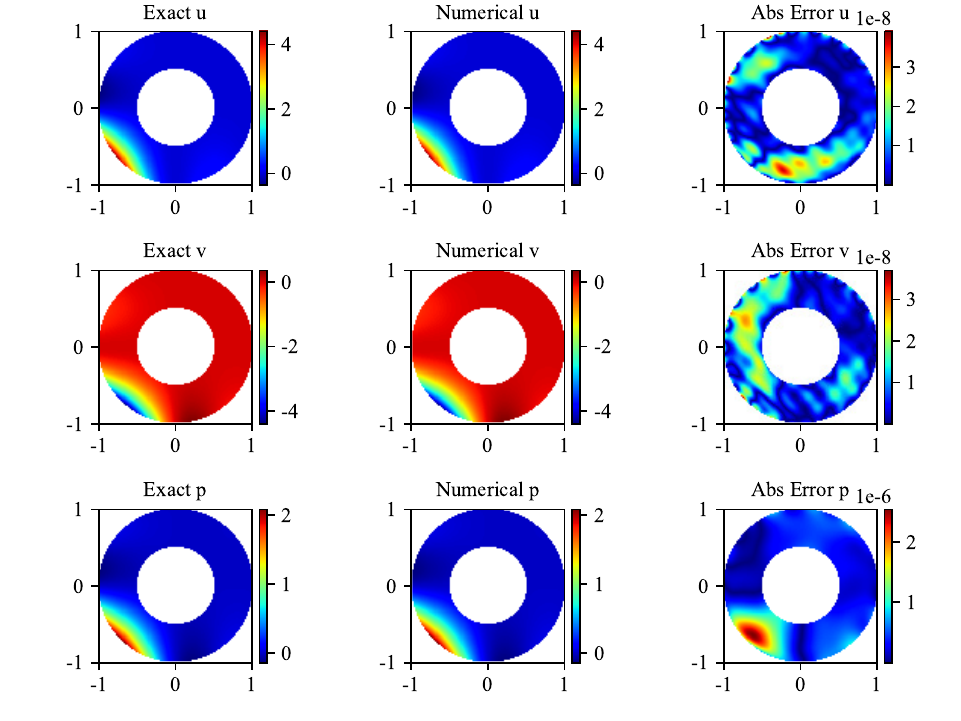}
	\caption{Velocity field, pressure field and absolute error, where $\ m=800,\ \gamma=100,\ N_c=5460,\ r_1=r_2=1$ and $r_3=0.5$ in Case 2 in Example \ref{ex_2d_mhd}.}
	\label{mhd-2d-case2}
\end{figure}

In this case, numerical integration is performed using the mean square error \eqref{MC_int}, with the number of integration points \(N_{\rm int}\) set to 10,000. The exact solutions, numerical solutions, and absolute errors of \(\mathbf{u}\) and \(p\) are shown in Figure \ref{mhd-2d-case2}.  

\begin{equation} \label{MC_int}
    \|v - v_{\rho}\|_0 = \sqrt{\frac{1}{N_{\rm int}} \sum_{i=1}^{N_{\rm int}} (v(\mathbf{x}_i) - v_{\rho}(\mathbf{x}_i))^2}.
\end{equation}

Table \ref{mhd-2d-non} provides the errors for different degrees of freedom when using the SP-RaNN-P method. Here, \(N_c = 5460\), \(r_1 = r_2 = 1\), \(r_3 = 0.5\), and \(\gamma = 100\). The results show high accuracy for the velocity \(\mathbf{u}\), magnetic field \(\mathbf{B}\) and pressure \(p\), with errors decreasing as the DoFs increases. Figure \ref{mhd-2d-case2} shows the velocity field, pressure field and error distribution for the case where \(m = 800\), \(\gamma = 100\), \(N_c = 5460\), \(r_1 = r_2 = 1\), and \(r_3 = 0.5\). The visualizations further illustrate the excellent performance of the proposed method. This example demonstrates the effectiveness of the SP-RaNN-P method in handling non‑simply‑connected domains. 

\begin{table}[H]
\centering
\begin{tabular}{ccccccccc}
\toprule
Method&DoFs&$e_{\mathbf{u}}$&$e_{\nabla \mathbf{u}}$&$\Vert \nabla \cdot \mathbf{u} \Vert_0$&$e_{\mathbf{B}}$&$e_{\nabla \mathbf{B}}$&$\Vert \nabla \cdot \mathbf{B} \Vert_0$&$e_{p}$ \\
\midrule
\multirow{5}{*}{SP-RaNN-P}&$300$&7.89E-05	&4.60E-03	&2.94E-12	&1.38E-04	&1.62E-04	&2.88E-12	&1.70E-04\\
&$600$&3.46E-07	    &6.40E-07	&1.40E-12	&6.83E-07	&5.45E-07	&1.75E-12	&2.80E-05\\
&$1200$&6.03E-08	&9.54E-08	&4.56E-13	&5.91E-07	&2.97E-07	&5.90E-13	&4.27E-06\\
&$2400$&2.31E-08	&3.78E-08	&2.66E-13	&6.79E-07	&3.13E-07	&3.41E-13	&1.93E-06\\
\bottomrule
\end{tabular}
\caption{Errors for different $m$ using SP-RaNN-P, where $N_c=5460,\ r_1=r_2=1,\ r_3=0.5$ and $\gamma=100$ in Example \ref{ex_2d_mhd}.}
\label{mhd-2d-non}
\end{table}

\begin{itemize}
\item Case 3: L-shape domain
\end{itemize}
We consider an L-shaped domain $\Omega = (-1,1)^2 \setminus \big((0,1) \times (-1,0)\big)$ with $H_a = N = R_m = 1$. The exact solution is given by
\begin{equation}
\left\{
\begin{array}{l}
    u =  x^2 (x-1)^2 y (y-1) (2y-1),\quad 
    v = y^2 (y-1)^2 x (x-1) (2x-1),  \\[3pt]
    B_1 =  \dfrac{5}{3} r^{\frac{2}{3}} \sin\!\left(\dfrac{2}{3}\theta\right),\quad 
    B_2 =  \dfrac{5}{3} r^{\frac{2}{3}} \cos\!\left(\dfrac{2}{3}\theta\right), \\[3pt]
    p = x(x-1)\!\left(x-\dfrac{1}{2}\right)y(y-1)\!\left(y-\dfrac{1}{2}\right).
\end{array}
\right.
\nonumber
\end{equation}

For this configuration, conventional FEM approaches employing nodal finite elements may produce spurious solutions. In contrast, our proposed method remains stable and free from such artifacts. However, due to the reduced regularity of the analytical solution near the re-entrant corner, the numerical accuracy of the strong-form formulation slightly deteriorates. To further enhance accuracy, one may adopt the weak formulations proposed in \cite{Shang2023} and \cite{Sun2024}.  

\begin{figure}[!htbp] 		
	\centering
	\includegraphics[scale=0.75]{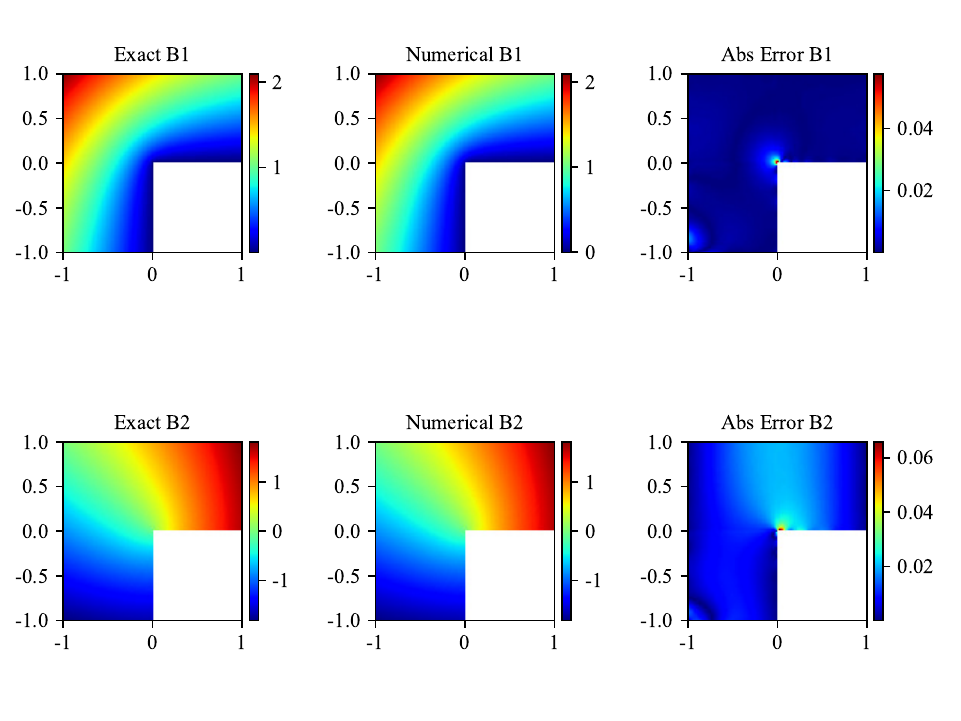}
	\caption{Magnetic field components and absolute errors, where $\ m=400,\ \gamma=100,\ N_c=5200,\ r_{\mathbf{u}}=1,\ r_{\mathbf{B}}=4,\ r_p=0.5$ in Case 3 in Example \ref{ex_2d_mhd}.}
	\label{mhd-2d-case3}
\end{figure}

To address this problem, we adopt a domain decomposition strategy following \cite{Dong2021}, dividing the domain $\Omega$ into three non-overlapping subdomains:
\[
\Omega_1 = (0,1)^2,\quad 
\Omega_2 = (-1,0)\times(0,1),\quad 
\Omega_3 = (-1,0)^2.
\]
In each subdomain, two SP-RaNNs, $\mathbf{u}^i_\rho$ and $\mathbf{B}^i_\rho$, are employed to approximate the local velocity and magnetic field, $\mathbf{u}^i$ and $\mathbf{B}^i$ ($i=1,2,3$), respectively.  
Continuity across interfaces is enforced by imposing $C^1$-continuity conditions along the shared boundaries of the subdomains. A single global RaNN, $p_\rho$, is used to approximate the pressure $p$ over the entire domain $\Omega$.  The network parameters are set as $r_{\mathbf{u}}=1$ and $r_{\mathbf{B}}=4$ for $\mathbf{u}^i_\rho$ and $\mathbf{B}^i_\rho$, respectively ($i=1,2,3$), with $r_p=0.5$ and $m=400$ neurons for each network. A total of eight Picard iterations are performed. The resulting magnetic field distributions and absolute errors are presented in Figure~\ref{mhd-2d-case3}.

\begin{example}[Unsteady Three-Dimensional MHD Equations]\label{ex_3d_mhd}

This example focuses on solving the unsteady three-dimensional MHD equations \eqref{MHD-model2}. In \cite{Zheng2018}, a fully divergence-free finite element method was introduced to solve these equations. We test the same case as Example 5.2 in \cite{Zheng2018}, where ${\rm Re}=\sigma=\mu=1$, with the domain $\Omega=(0,1)^3$ and time interval $I=(0,5)$. The analytic solutions are:
$$
\mathbf{u}=({\rm sin}(t){\rm sin}(y),0,0)^{\rm T},\ \mathbf{B}=(0,0,\rm cos(t+x))^T,\ p=x+y+z-1.5.
$$
\end{example}

To evaluate the performance of SP-RaNN, we compare it against the fully divergence-free FEM method from \cite{Zheng2018} over the time interval $I = (0,0.2)$. Two SP-RaNNs are employed to approximate $\mathbf{u}$ and $\mathbf{B}$, while a separate RaNN is used for $p$, with parameters set to $r_1 = r_2 = r_3 = 0.1$ and $N_c = 3072$ (comprising $N_1 = 2400$ interior points and $N_2 = 672$ boundary points). We apply the SP-RaNN method using three nonlinear iteration steps, either through the Picard or Newton method. Table \ref{compare-mhd-3d} presents the numerical errors of SP-RaNN at $t = 0.2$, while Table \ref{fully_FEM} provides the numerical errors of the fully divergence-free FEM method at $t = 0.2$, utilizing a tetrahedral mesh. In this approach, $\mathbf{u}$ is approximated using $H({\rm div}, \Omega)$-conforming finite elements, $p$ is represented by fully discontinuous finite elements, and the magnetic potential $\mathbf{A}$ is approximated using $H({\rm curl}, \Omega)$-conforming edge elements, with the magnetic field computed as $\mathbf{B}_h = {\rm curl} \, \mathbf{A}_h$. It is important to note that SP-RaNN employs a space-time approach, whereas the fully divergence-free FEM method uses a Crank-Nicolson scheme. Additionally, \cite{Zheng2018} implements a linearly extrapolated scheme to handle nonlinear terms. From the results in Tables \ref{compare-mhd-3d} and \ref{fully_FEM}, it is evident that SP-RaNN achieves higher accuracy with fewer degrees of freedom compared to the FEM method.

\begin{table}[H]
    \centering
    \begin{tabular}{cccccccccc}
    \toprule
    Method&DoFs&$\Vert \mathbf{u}-\mathbf{u}_{\rho}\Vert_0$&$|\mathbf{u}-\mathbf{u}_{\rho}|_1$&$\Vert p-p_{\rho}\Vert_0$&$\Vert \mathbf{B}-\mathbf{B}_{\rho}\Vert_0$&$\Vert \nabla \cdot \mathbf{u}_{\rho} \Vert_0$&$\Vert \nabla \cdot \mathbf{B}_{\rho} \Vert_0$& time (s) \\
    \midrule
    \multirow{5}{*}{Picard}&$300$&3.73E-04	&3.61E-03	&4.19E-03	&5.07E-04	&1.31E-14	&2.16E-14&0.84\\
    &$600$&1.55E-04	    &1.31E-03	&1.79E-03	&2.04E-04	&2.51E-13	&4.03E-13    &1.48\\
    &$1200$&4.10E-06	&3.66E-05	&8.35E-05	&3.56E-06	&7.84E-13	&1.03E-12   &3.39\\
    &$2400$&8.93E-07	&8.75E-06	&1.96E-05	&6.76E-07	&1.54E-12	&2.47E-12   &10.72\\
    &$4800$&5.31E-07	&4.79E-06	&1.12E-05	&4.06E-07	&1.13E-12	&1.88E-12   &51.24\\
    \midrule
    \multirow{5}{*}{Newton}&$300$&3.88E-04	&3.49E-03	&4.46E-03	&5.02E-04	&1.55E-14	&2.45E-14	&1.05 \\
    &$600$&1.78E-04	    &1.43E-03	&2.15E-03	&2.28E-04	&2.47E-13	&4.22E-13	&1.82 \\
    &$1200$&3.84E-06	&3.47E-05	&8.14E-05	&3.33E-06	&7.49E-13	&1.25E-12	&4.01 \\
    &$2400$&9.28E-07	&8.99E-06	&2.04E-05	&8.91E-07	&1.55E-12	&2.67E-12	&12.84 \\
    &$4800$&4.02E-07	&3.81E-06	&8.91E-06	&3.69E-07	&1.11E-12	&1.82E-12	&55.78 \\
    \bottomrule
    \end{tabular}
    \caption{Errors of SP-RaNN at $t=0.2$ and CPU time, where $N_c=3072,\ r_1=r_2=r_3=0.1$ and $\gamma=100$ in Example \ref{ex_3d_mhd}.}
    \label{compare-mhd-3d}
\end{table}

\begin{table}[H]
    \centering
    \begin{tabular}{ccccccccc}
    \toprule
&$(\tau, h)$&DoFs&$\Vert \mathbf{u}-\mathbf{u}_{h}\Vert_0$&$|\mathbf{u}-\mathbf{u}_{h}|_{1,h}$&$\Vert p-p_{h}\Vert_0$&$\Vert \mathbf{B}-\mathbf{B}_{h}\Vert_0$&$\Vert \nabla \cdot \mathbf{u}_{h} \Vert_0$&$\Vert \nabla \cdot \mathbf{B}_{h} \Vert_0$ \\
    \midrule
\multirow{5}{*}{k=1}&$(\tau_0,h_0)$&604&3.87E-04	&1.29E-02	&1.77E-01	 &7.50E-02   &1.71E-09&0\\
&$(\tau_0,h_0)/2$&4184               &1.02E-04	&6.23E-03	&8.86E-02	 &3.69E-02   &5.28E-10&0\\
&$(\tau_0,h_0)/4$&31024              &2.83E-05	&3.04E-03	&4.43E-02	 &1.83E-02   &7.27E-11&0\\
&$(\tau_0,h_0)/8$&238688             &7.84E-06	&1.50E-03	&2.21E-02	 &9.08E-03   &5.18E-10&0\\
\midrule
\multirow{5}{*}{k=2}&$(\tau_0,h_0)$&1062&3.71E-04	&1.28E-02	&1.77E-01	&5.14E-03     &1.19E-09  &4.43E-16\\
&$(\tau_0,h_0)/2$&7380                &9.61E-05	&6.19E-03	&8.84E-02	&1.26E-03     &3.40E-10  &1.33E-15\\
&$(\tau_0,h_0)/4$&54792               &2.69E-05	&3.02E-03	&4.42E-02	&3.12E-04     &1.23E-10  &5.42E-15\\
&$(\tau_0,h_0)/8$&421776              &7.49E-06	&1.48E-03	&2.21E-02	&7.75E-05     &7.61E-10  &2.12E-14\\
    \bottomrule
    \end{tabular}
    \caption{
    Errors of the fully divergence-free FEM method (\cite{Zheng2018}) at $t = 0.2$ for varying time steps $\tau$ and mesh sizes $h$. Here, $k$ denotes the order of the piecewise polynomials used to approximate $\mathbf{A}$ ($\mathbf{B}_h = {\rm curl} \, \mathbf{A}_h$), while 1st-order elements and piecewise constant functions are used to approximate $\mathbf{u}$ and $p$, respectively.
    }
    \label{fully_FEM}
\end{table}

We also assess the performance of SP-RaNN over a longer time interval, $I = (0,5)$. As before, two SP-RaNNs are used to approximate $\mathbf{u}$ and $\mathbf{B}$, with a separate RaNN for $p$, using parameters $r_1 = r_2 = r_3 = 0.3$ and $N_c = 3072$. The space-time $L^2$ and semi-$H^1$ errors for SP-RaNN are displayed in Table \ref{mhd-3d-T}, demonstrating the method's efficiency and accuracy in long-time simulations.

\begin{table}[H]
    \centering
    \begin{tabular}{cccccccccc}
    \toprule
    Method&DoFs&$\Vert \mathbf{u}-\mathbf{u}_{\rho}\Vert_0^{st}$&$|\mathbf{u}-\mathbf{u}_{\rho}|_1^{st}$&$\Vert p-p_{\rho}\Vert_0^{st}$&$\Vert \mathbf{B}-\mathbf{B}_{\rho}\Vert_0^{st}$&$\Vert \nabla \cdot \mathbf{u}_{\rho} \Vert_0^{st}$&$\Vert \nabla \cdot \mathbf{B}_{\rho} \Vert_0^{st}$& time (s) \\
    \midrule
    \multirow{5}{*}{Picard}&$300$&7.51E-02	&3.25E-01	&2.81E-01	&1.32E-01	&4.23E-15	&6.91E-15	&0.83 \\
    &$600$&1.52E-02	    &9.88E-02	&9.87E-02	&2.94E-02	&1.29E-14	&1.97E-14	&1.57 \\
    &$1200$&2.13E-03	&1.48E-02	&1.56E-02	&4.17E-03	&2.37E-14	&3.89E-14	&3.49 \\
    &$2400$&3.30E-04	&2.50E-03	&2.94E-03	&7.42E-04	&7.77E-14	&1.28E-13	&11.45 \\
    &$4800$&3.96E-05	&3.50E-04	&3.40E-04	&1.39E-04	&2.76E-13	&4.66E-13	&52.64 \\
    \midrule
    \multirow{5}{*}{Newton}&$300$&7.29E-02	&3.39E-01	&2.57E-01	&1.24E-01	&4.75E-15	&6.58E-15	&1.08 \\
    &$600$&1.55E-02	    &9.72E-02	&9.37E-02	&3.21E-02	&1.32E-14	&2.00E-14	&1.87 \\
    &$1200$&2.12E-03	&1.43E-02	&1.50E-02	&4.87E-03	&2.44E-14	&4.04E-14	&4.09 \\
    &$2400$&3.30E-04	&2.54E-03	&2.89E-03	&7.77E-04	&7.59E-14	&1.26E-13	&12.46 \\ 
    &$4800$&5.18E-05	&4.28E-04	&4.48E-04	&1.53E-04	&2.60E-13	&5.09E-13	&57.26 \\
    \bottomrule
    \end{tabular}
    \caption{Space-time errors over $(0,1)^3\times(0,5)$ and CPU time for SP-RaNN, where $N_c=3072,\ r_1=r_2=r_3=0.3$ and $\gamma=100$ in Example \ref{ex_3d_mhd}.}
    \label{mhd-3d-T}
\end{table}

\section{Summary}
\label{sec5}

By embedding the divergence-free condition directly into the structure of randomized neural networks, we propose a Structure-Preserving Randomized Neural Network method for solving incompressible MHD problems. This special design ensures that the SP-RaNN inherently satisfies the divergence-free condition pointwise, preserving the physical and mathematical structure of the problem.

The SP-RaNN method offers several significant advantages due to its structure-preserving nature:
\begin{description}
\item[Automatically divergence-free: ] The method eliminates the need for additional constraints, ensuring physical consistency and higher accuracy with fewer degrees of freedom.

\item[Efficient optimization: ] Since the training process only involves solving linear optimization problems, the SP-RaNN reduces optimization errors and computational costs compared to standard neural network approaches.

\item[Reduction of error accumulation:] Using a space-time framework, the SP-RaNN avoids error accumulation typically introduced by iterative time-stepping schemes.

\item[Robustness:] The SP-RaNN remains stable and effective for problems with high Reynolds and Hartmann numbers. 
\end{description}

Future research directions include:
\begin{enumerate}
\item Combining the adaptive growing RaNN framework proposed in \cite{Dang2024} with SP-RaNN to further enhance efficiency.
\item  Conducting theoretical analysis of the convergence rate to establish rigorous performance guarantees.
\item  Designing other structure-preserving neural networks tailored to specific physical problems, ensuring mathematical consistency across diverse applications. 
\end{enumerate}

This structure-preserving approach not only advances the accuracy and efficiency of numerical solutions but also establishes a foundation for solving complex physical problems while respecting their inherent mathematical structures.


\begin{thebibliography}{99}
    \bibitem{Barron1993}
    A. R. Barron, Universal approximation bounds for superpositions of a sigmoidal function, \emph{IEEE Transactions on Information Theory} {\bf 39} (1993), 930-945.
    
    \bibitem{Chen2022}
    J. Chen, X. Chi, W. E and Z. Yang, Bridging Traditional and Machine Learning-based Algorithms for Solving PDEs: The Random Feature Method, \emph{Journal of Machine Learning} {\bf 1} (2022), 268-298.
    
    \bibitem{Chen1995}
    T. Chen and H. Chen, Universal approximation to nonlinear operators by neural networks with arbitrary activation functions and its application to dynamical systems, \emph{IEEE Transactions on Neural Networks} {\bf 6} (1995), 911–917.
    
    \bibitem{Cockburn2007}
    B. Cockburn, G. Kanschat and D. Sch\"otzau, A note on discontinuous Galerkin divergence-free solutions of the Navier–Stokes equations, \emph{Journal of Scientific Computing} {\bf 31} (2007), 61-73.
    
    \bibitem{Cockburn2004}
    B. Cockburn, F. Li and C. Shu, Locally divergence-free discontinuous Galerkin methods for the Maxwell equations, \emph{Journal of Computational Physics} {\bf 194} (2004), 588-610.
    
    \bibitem{Dang2023}
    H. Dang and F. Wang, Local Randomized Neural Networks with Hybridized Discontinuous Petrov-Galerkin Methods for Stokes-Darcy Flows, \emph{Physics of Fluids} {\bf 36} (2024), 087138.
    
    \bibitem{Dang2024}
    H. Dang, F. Wang and S. Jiang, Adaptive Growing Randomized Neural Networks for Solving Partial Differential Equations, (2024), arXiv:2408.17225v2

    \bibitem{Davidson2001}
    P. A. Davidson, An Introduction to Magnetohydrodynamics, Cambridge: Cambridge University Press, 2001.
    
    \bibitem{Dong2021}
    S. Dong and Z. Li, Local extreme learning machines and domain decomposition for solving linear and nonlinear partial differential equations, \emph{Computer Methods in Applied Mechanics and Engineering} {\bf 387} (2021), 114129.
    
    \bibitem{Dong2023}
    S. Dong and Y. Wang. A method for computing inverse parametric PDE problems with random-weight neural networks. \emph{Journal of Computational Physics} {\bf 489} (2023), 112263.
    
    \bibitem{Dong2018}
    X. Dong and Y. He, Optimal convergence analysis of Crank–Nicolson extrapolation scheme for the three-dimensional incompressible magnetohydrodynamics, \emph{Computers \& Mathematics with Applications} {\bf 76} (2018), 2678–2700.
    
    \bibitem{E2017}
    W. E and B. Yu, The deep Ritz method: A deep learning-based numerical algorithm for solving variational problems, \emph{Communications in Mathematics and Statistics} {\bf 6} (2017), 6–12.
    
    \bibitem{Gao2019}
    H. Gao and W. Qiu. A semi-implicit energy conserving finite element method for the dynamical incompressible magnetohydrodynamics equations, \emph{Computer Methods in Applied Mechanics and Engineering} {\bf 346} (2019), 982-1001.
    
    \bibitem{Gerbeau2000}
    J. F. Gerbeau, A stabilized finite element method for the incompressible magnetohydrodynamic equations, \emph{Numerische Mathematik} {\bf 87} (2000), 83-111.
    
    \bibitem{Gerbeau2006}
    J. F. Gerbeau, C. L. Bris, and T. Lelièvre, Mathematical Methods for the Magnetohydrodynamics of Liquid Metals, Numerical Mathematics and Scientific Computation, Oxford University Press, New York, 2006.
    
    \bibitem{Ghia1982}
    U. Ghia, K.N. Ghia and C.T. Shin, High-Re solutions for incompressible flow using the Navier-Stokes equations and a multigrid method, \emph{Journal of Computational Physics} {\bf 48} (1982), 387-411.

    \bibitem{Greif2010}
    C. Greif, D. Li, D. Sch\"otzau and X. Wei, A mixed finite element method with exactly divergence-free velocities for incompressible magnetohydrodynamics, \emph{Computer Methods in Applied Mechanics and Engineering} {\bf 199} (2010), 2840-2855.
    
    \bibitem{Guermond2003}
    J. L. Guermond and P. D. Minev, Mixed finite element approximation of an MHD problem involving conducting and insulating regions: the 3D case, \emph{Numerical Methods for Partial Differential Equations} {\bf 19} (2003), 709-731.
    
    \bibitem{Gunzburger1991}
    M.D. Gunzburger, A.J. Meir, and J.S. Peterson, On the existence and uniqueness and finite element approximation of solutions of the equations of stationary incompressible magnetohydrodynamics, \emph{Mathematics of Computation} {\bf 56} (1991), 523-563.
    
    \bibitem{Zheng2018}
    R. Hiptmair, L. Li, S. Mao, and W. Zheng, A fully divergence-free finite element method for magnetohydrodynamic equations, \emph{Mathematical Models and Methods in Applied Sciences} {\bf 28} (2018), 659-695.
    
    \bibitem{Hiptmair2007}
  R. Hiptmair and J. Xu, Nodal Auxiliary Space Preconditioning in H(curl) and H(div) Spaces, \emph{SIAM Journal on Numerical Analysis} {\bf 45} (2007), 2483-2509.

    \bibitem{Hu2015}
    Kaibo Hu and Jinchao Xu, Structure-preserving finite element methods for stationary MHD models, \emph{Mathematics of Computation} {\bf 88} (2015), 553-581.

    \bibitem{Hu2017}
    Kaibo Hu, Yicong Ma and Jinchao Xu, Stable finite element methods preserving $\nabla \cdot \mathbf{B}=0$ exactly for MHD models, \emph{Numerische Mathematik} {\bf 135} (2017), 371-396.

    \bibitem{Huang2011}
    G. Huang, D. Wang and Y. Lan, Extreme learning machines: A survey, \emph{International Journal of Machine Learning and Cybernetics} {\bf 2} (2011), 107–122.
    
    \bibitem{Huang2006}
    G. Huang, Q. Zhu and C. K. Siew, Extreme learning machine: theory and applications, \emph{Neurocomputing} {\bf 70} (2006), 489–501.
    
    \bibitem{Huang2012}
    J. Huang and S. Zhang, A divergence-free finite element method for a type of 3D Maxwell equations, \emph{Applied Numerical Mathematics} {\bf 62} (2012) 802-813.
    
    \bibitem{Igelnik1995}
    B. Igelnik and Y.H. Pao, Stochastic choice of basis functions in adaptive function approximation and the functional-link net, \emph{IEEE Transactions on Neural Networks} {\bf 6} (1995), 1320–1329.
    
    \bibitem{Jardin2010}
    S. Jardin, Computational Methods in Plasma Physics, CRC Press, 2010.
    
    \bibitem{Jiang1996}
    B-N. Jiang, J. Wu and LA. Povinelli, The origin of spurious solutions in computational electromagnetics, \emph{Journal of Computational Physics} {\bf 125} (1996), 104-123.
    
    \bibitem{George2021}
    X. Jin, S. Cai, H. Li and G. E. Karniadakis, NSFnets (Navier-Stokes flow nets): Physics-informed neural networks for the incompressible Navier-Stokes equations, \emph{Journal of Computational Physics} {\bf 426} (2021), 109951.
    
    \bibitem{John2017}
    V. John, A. Linke, C. Merdon, M. Neilan, and L. G. Rebholz, On the Divergence Constraint in Mixed Finite Element Methods for Incompressible Flows, \emph{SIAM Review} {\bf 59} (2017), 492-544.
    
    \bibitem{Zheng2021}
    L. Li, D. Zhang and W. Zheng, A constrained transport divergence-free finite element method for incompressible MHD equations, \emph{Journal of Computational Physics} {\bf 428} (2021), 109980.
    
    \bibitem{Li2017}
     L. Li and W. Zheng, A robust solver for the finite element approximation of stationary incompressible MHD equations in 3D, \emph{Journal of Computational Physics} {\bf 351} (2017), 254-270.
    
    \bibitem{Li2023}
    Y. Li and F. Wang, Local randomized neural networks with finite difference methods for interface problems, \emph{Journal of Computational Physics} {\bf 529} (2025), 113847.
    
    \bibitem{Linke2009}
    Linke A, Collision in a cross-shaped domain-a steady 2d Navier–Stokes example demonstrating the importance of mass conservation in CFD, \emph{Computer Methods in Applied Mechanics and Engineering} {\bf 198} (2009), 3278-3286.
    
    \bibitem{Linke2014}
    A. Linke, On the role of the Helmholtz decomposition in mixed methods for incompressible flows and a new variational crime, \emph{Computer Methods in Applied Mechanics and Engineering} {\bf 268} (2014), 782-800.
    
    \bibitem{Muller2001}
    U. M\"uller and L. B\"uhler, Magnetofluiddynamics in Channels and Containers, Springer-Verlag, Berlin, 2001.
    
    \bibitem{Neufeld2023}
    A. Neufeld and P. Schmocker, Universal Approximation Property of Random Neural Networks, arXiv:2312.08410.
    
    \bibitem{Pao1994}
    Y. Pao, G. Park and D. Sobajic, Learning and generalization characteristics of the random vector functional-link net, \emph{Neurocomputing} {\bf 6} (1994), 163–180.
    
    \bibitem{Pozrikidis1997}
    C. Pozrikidis, Introduction to Theoretical and Computational Fluid Dynamics, Oxford University Press, 1997.

    \bibitem{George2019}
    M. Raissi, P. Perdikaris and G. E. Karniadakis, Physics-informed neural networks: A deep learning framework for solving forward and inverse problems involving nonlinear partial differential equations, \emph{Journal of Computational Physics} {\bf 378} (2019), 686–707.
    
    \bibitem{Schotzau2004}
    D. Sch\"otzau, Mixed finite element methods for stationary incompressible magnetohydrodynamics, \emph{Numerische Mathematik} {\bf 96} (2004), 771-800.
    
    \bibitem{Shang2023}
    Y. Shang, F. Wang and J. Sun, Randomized neural network with petrov–galerkin methods for solving linear and nonlinear partial differential equations, \emph{Communications in Nonlinear Science and Numerical Simulation} {\bf 127} (2023), 107518.
    
    \bibitem{Shang2024}
    Y. Shang and F. Wang, Randomized Neural Networks with Petrov–Galerkin Methods for Solving Linear Elasticity and Navier–Stokes Equations, \emph{Journal of Engineering Mechanics} {\bf 150} (2024), 04024010.
    
    \bibitem{Sirignano2019}
    J. Sirignano and K. Spiliopoulos, DGM: A deep learning algorithm for solving partial differential equations, \emph{Journal of Computational Physics} {\bf 375} (2018), 1339–1364.
    
    \bibitem{Feng2019}
    H. Su, S. Mao and X. Feng, Optimal error estimates of penalty based iterative methods for steady incompressible magnetohydrodynamics equations with different viscosities, \emph{Journal of Scientific Computing} {\bf 79} (2019), 1078-1110.
    
    \bibitem{Sun2024}
    J. Sun, S. Dong and F. Wang, Local randomized neural networks with discontinuous galerkin methods for partial differential equations, \emph{Journal of Computational and Applied Mathematics} {\bf 445} (2024), 115830.
    
    \bibitem{Sun2024-2}
    J. Sun and F. Wang, Local randomized neural networks with discontinuous Galerkin methods for diffusive-viscous wave equation, \emph{Computers Mathematics with Applications} {\bf 154} (2024), 128-137.
    
    \bibitem{Toth2000}
    G. Tóth, The $\nabla \cdot \mathbf{B}=0$ Constraint in Shock-Capturing Magnetohydrodynamics Codes, \emph{Journal of Computational Physics} {\bf 161} (2000), 605-652.
    
    \bibitem{Xie2013}
    Z. Xie, B. Wang, and Z. Zhang, Space-Time Discontinuous Galerkin Method for Maxwell’s Equations, \emph{Communications in Computational Physics} {\bf 14} (2013), 916–939.
    
    \bibitem{Xu2010}
    X. Xu and S. Zhang, A new divergence-free interpolation operator with applications to the Darcy-Stokes-Brinkman equations, \emph{SIAM Journal of Scientific Computing} {\bf 32} (2010), 855-874.
    
    \bibitem{Zhang2015}
    G. Zhang, Y. He and Y. Zhang, Unconditional convergence of the Euler semi-implicit scheme for the three-dimensional incompressible MHD equations, \emph{IMA Journal of Numerical Analysis} {\bf 35} (2015), 767-801.

    \bibitem{Zhang2014}
    G. Zhang, Y. He and Y. Zhang, Streamline diffusion finite element method for stationary incompressible magnetohydrodynamics, \emph{Numerical Methods for Partial Differential Equations} {\bf 30} (2014), 1877-1901.
    
    \bibitem{Xie2024}
    M. Zhang, T. Zhang and X. Xie, Robust globally divergence-free Weak Galerkin finite element method for incompressible Magnetohydrodynamics flow, \emph{Communications in Nonlinear Science and Numerical Simulation} {\bf 131} (2024), 107810.
 
    \bibitem{Hou2015}
 Y. Zhang and Y. Hou, Numerical analysis of the Crank–Nicolson extrapolation time discrete scheme for magnetohydrodynamics flows, \emph{Numerical Methods for Partial Differential Equations} {\bf 31} (2015), 2169-2208.

\end{thebibliography}
\end{document}